\documentclass[aps,floats,twocolumn]{revtex4-1}
\usepackage{amsmath,amssymb}
\usepackage{graphicx,epsfig}
\usepackage[greek,english]{babel}
\usepackage{bbold}
 
\newcommand{\be}{\begin{equation}}
\newcommand{\ee}{\end{equation}}
\newcommand{\beqn}{\begin{eqnarray}}
\newcommand{\eeqn}{\end{eqnarray}}

\begin{document}
\bibliographystyle {plain}

\def\oppropto{\mathop{\propto}} 
\def\opsimeq{\mathop{\simeq}}
\def\opoverderline{\mathop{\overline}}
\def\operarrow{\mathop{\longrightarrow}}
\def\opsim{\mathop{\sim}}

\def\fig#1#2{\includegraphics[height=#1]{#2}}
\def\figx#1#2{\includegraphics[width=#1]{#2}}

\title{ Strong Disorder RG approach - a short review of recent developments  } 

\author{Ferenc Igl{\'o}i}
\email{igloi.ferenc@wigner.mta.hu}
\affiliation{Wigner Research Centre for Physics, Institute for Solid State Physics and Optics, H-1525 Budapest, P.O. Box 49, Hungary}
\affiliation{Institute of Theoretical Physics, Szeged University, H-6720 Szeged, Hungary}
\author{C\'ecile Monthus}
\email{cecile.monthus@cea.fr}
\affiliation{Institut de Physique Th\'eorique, Universit\'e Paris Saclay, CNRS, CEA, 91191 Gif-sur-Yvette, France}
\date{\today}

\begin{abstract}

\end{abstract}

\maketitle
\vskip 10cm
\clearpage
\tableofcontents
\clearpage


\section{ Introduction }

The Strong Disorder Renormalization Group (SDRG) approach has been introduced by Ma, Dasgupta and Hu \cite{ma79} and later further developed by D. Fisher \cite{fisher92} to study the low-energy excitations and spatial and temporal correlations of random quantum spin chains. In these systems, the critical properties are controlled by so called Infinite-Disorder-Fixed-Points (IDFPs), at which disorder fluctuations dominate over quantum fluctuations, so that the calculated properties (critical exponents and scaling functions) are asymptotically exact. Soon after Fisher's results, the RG approach has been applied to a large number of disordered systems, either quantum (in one and higher dimensions) or classical (Sinai random walk, classical random spin chains and polymers, stochastic systems with quenched disorder, etc.). For the models where exact results are available via other approaches,
the RG method has been not only able to reproduce them correctly, but has allowed in addition to predict many new critical exponents and scaling functions. All the developments that have occoured before 2005 have been already summarised in our Review \cite{igloi05review}. 
The goal of the present colloquium paper is thus to give an overview of the various developments since 2005.
We will stress the physical ideas and the SDRG-way-of-thinking for each type of problem,
but we will usually omit the detailed derivation of results that can be found in the original papers.

This colloquium paper is organized as follows. 
We begin with the ground state properties of random quantum systems, 
such as the random transverse-field Ising model in different dimensions $d$  in Section \ref{sec:QuantumIsing},
the effects of Long-Ranged interactions in Section \ref{sec:LR},
as well as various other quantum models (antiferromagnets, Ashkin-Teller models, anyonic models, the superfluid-insulator transition) 
 in Section \ref{sec:other}. 
The scaling properties of the entanglement in critical ground states are described in Section \ref{sec:entanglement}.
We then focus on the dynamical properties of Localized and Many-Body-Localized models
within three different settings, namely the unitary dynamics of isolated models in Section \ref{sec:MBL},
 the Floquet dynamics of periodically driven models in Section \ref{sec:Floquet},
and the dissipative dynamics of open quantum models in Section \ref{sec:open}.
Section \ref{sec_AndersonLoc} is then devoted to Anderson Localization tight-binding models for electrons. 
We then turn to classical disordered models,
with the random contact process for epidemic spreading in section \ref{sec:non-equilibrium classical},
the renormalization of general master equations with randomness in section \ref{sec:mastereq},
the dynamics of random classical oscillators with or without dissipation in section \ref{sec:oscillators},
as well as some other miscellaneous topics in section \ref{sec:otherclassical}.

\section{ Random Quantum Ising Model in various dimensions $d$  }

\label{sec:QuantumIsing}

The quantum Ising model with random couplings and/or with random transverse fields (RTIM) is the prototype of
disordered quantum magnets having discrete symmetry. The model is defined by the Hamiltonian:
\be
{\cal H} =
-\frac{1}{2}\sum_{ij} J_{ij}\sigma_i^x \sigma_{j}^x-\frac{1}{2}\sum_{i} h_i \sigma_i^z\;,
\label{eq:H}
\ee
where the $\sigma_i^{x,z}$ are Pauli-matrices and $i$, $j$ denote sites of a lattice. In Eq.(\ref{eq:H}) the couplings $J_{ij}$
and the transverse fields $h_i$ are independent random variables, which are taken from the distributions, $p(J)$ and $q(h)$, respectively. In the following, we will discuss the case of ferromagnetic models $J_{ij} \geq 0$,
where the order parameter is the magnetization,
but it should be stressed that this is not restrictive : 
the SDRG approach has been applied as well to the spin-glass case where the sign of the couplings is also random.
The numerical computations are usually done either with two box-distributions
$p(J)=\Theta(J)\Theta(1-J)$ and $q(h)=\dfrac{1}{h_b}\Theta(h)\Theta(h_b-h)$ ($\Theta(x)$ being the Heaviside step-function),
or with the fixed-h distribution $q(h)=\delta(h-h_f)$, while $p(J)$ follows the box-distribution as before. The quantum control parameter is defined as $\theta=\overline{\ln h}-\overline{\ln J}$, where $\overline{x}$ stands for the average value of the variable $x$ over quench disorder. In any dimension the RTIM has a zero-temperature quantum phase transition, located at some $\theta_c$, separating a paramagnetic disordered ($\theta>\theta_c$) and a ferromagnetic ordered phase ($\theta<\theta_c$).

The simplest geometry that one can consider is the one-dimensional chain with nearest-neighbour couplings, which has been solved analytically through the Strong Disorder RG method by Fisher \cite{fisher92}, as described in details in Review \cite{igloi05review}. Later, more general geometries have been considered, such as ladders; star-like objects; 
hyper-cubic lattices in dimensions $d=2,3,4$ ; Bethe lattices, Erd\H os-R\'enyi random graphs and other type of complex networks, as well as systems with long-range interactions. Besides the bulk quantities, the critical behaviour of several observables at surfaces, edges, wedges and cones have also been calculated.

\subsection{Strong Disorder RG rules}

%
%

The basic steps of the renormalization procedure are identical in every geometry: at each step the largest term of the Hamiltonian in Eq.(\ref{eq:H}), denoted by $\Omega$, is eliminated and new terms are generated through a second-order perturbation calculation between the remaining degrees of freedom.\\
i) If the largest term is a coupling, say $J_{ij}=\Omega$, then the two connected sites, $i$ and $j$ are coupled to form a cluster. This spin-cluster then perceives an effective transverse field of strength
\be
\tilde{h}_{ij}\approx \frac{h_i h_j}{J_{ij}}\;,
\label{h_tilde}
\ee
 and the magnetic moments transform additively: $\mu_{ij}=\mu_i + \mu_j$, while originally $\mu_i = \mu_j=1$.\\
ii) If the largest term is a transverse field, say $h_i=\Omega$, then this site has negligible contribution to the (longitudinal) susceptibility, and is therefore decimated out. At the same time, new effective couplings are generated between all sites, say $j$ and $k$, which were nearest neighbours of $i$. The new contributions to the couplings are given by: $\tilde{J}_{jk}\approx \frac{J_{ij} J_{ik}}{h_{i}}$.\\ In higher dimensions $d>1$, the topology of the lattice is modified during the renormalization and often the new contribution $\tilde{J}_{jk}\approx \frac{J_{ij} J_{ik}}{h_{i}}$ concerns a pair of sites that were already coupled via some coupling $J_{jk}$. In this case, one can use their sum (\textit{sum rule}), or their maximum value (\textit{maximum rule}) as the renormalized coupling
\be
\tilde{J}_{jk}\approx {\rm max}\left[J_{jk}, \frac{J_{ij} J_{ik}}{h_{i}}\right]\;.
\label{J_tilde}
\ee
The maximum rule is justified at an IDFP, furthermore the numerical algorithms are more efficient in this case,
as explained in Refs.\onlinecite{kovacs09,kovacs10,kovacs11a,kovacs11b}.
 If, however, the critical behaviour is controlled by a conventional random fixed point, 
such as in the superfluid-insulator transition (see section \ref{sec:superfluid-insulator}), 
the sum rule can lead to more accurate results \cite{iyer12}.

\subsection{Solution in one dimension - a reminder\cite{igloi05review}}

In one dimension with nearest neighbour couplings, the topology does not change during renormalization. The cut-off ($\Omega$) dependence of the distribution functions (couplings, transverse fields, lengths, moments, etc) is written in the form of integro-differential equations, which are solved analytically. At the fixed point $\Omega \to 0$, the distribution of transverse fields and that of the couplings are given in the form:
\begin{eqnarray}
P_0(h,\Omega)&=&\frac{p_0(\Omega)}{\Omega}\left(\frac{\Omega}{h}\right)^{1-p_0(\Omega)}
\label{Psol}\\
R_0(J,\Omega)&=&\frac{r_0(\Omega)}{\Omega}\left(\frac{\Omega}{J}\right)^{1-r_0(\Omega)}
\label{Rsol}
\;,
\end{eqnarray}
\subsubsection{Critical point - infinite disorder fixed point}

At the critical point where the decimation of couplings and transverse fields are symmetric, we have 
\begin{equation}
p_0=r_0=\frac{1}{\ln(\Omega_0/\Omega)}\;,
\label{sol0}
\end{equation}
where $\Omega_0$ is a reference energy scale. This is an infinite disorder fixed point, since the ratio of typical couplings and transverse fields is going to zero or to infinity. Therefore the decimation steps are asymptotically correct and consequently the calculated critical properties are asymptotically exact. At this fixed point the relation between length-scale, $L$, and energy-scale, $\Omega$ is given by the activated scaling
\be
\ln \left(\frac{\Omega_0}{\Omega}\right) \sim L^{\psi},\quad \psi=1/2\;,
\label{Psi}
\ee
and the average moment of clusters scales as:
\begin{equation}
\overline{\mu}= \overline{\mu}_0 \left[ \ln
\left(\frac{\Omega_0}{\Omega}\right)\right]^{\phi},~~
\phi=\frac{1+\sqrt{5}}{2}\;.
\label{Phi}
\end{equation}
leading to the fractal dimension $d_f$ 
\begin{equation}
\overline{\mu} \sim L^{d_f},\quad d_f=\psi \phi=\frac{1+\sqrt{5}}{4}\;.
\label{d_f}
\end{equation}

\subsubsection{Griffiths phases - strong disorder fixed points}

Outside the critical point, the decimation of the couplings and the transverse fields is asymmetric, which is characterized by the parameter $\Delta=(p_0-r_0)/2$. $\Delta$ is expressed in terms of the original distributions as:
\begin{equation}
\left[\left(\frac{J^2}{h^2}\right)^{\Delta}\right]_{\rm av}=1
\;,
\label{zeq}
\end{equation}
and close to the critical point: $\Delta \sim \theta$, since $\theta_c=0$.
In the paramagnetic or disordered Griffiths phase,
 almost exclusively transverse fields are decimated out and the solution close to the line of fixed points, i.e. in the limit $\Omega/\Omega_0 \ll 1$ is given by:
\begin{equation}
p_0=2 \Delta,\quad r_0 \sim \left({\Omega / \Omega_0} \right)^{2 \Delta}\;.
\label{sol1}
\end{equation}
In the ferromagnetic or ordered Griffiths phase the expressions for $p_0$ and $r_0$ are reversed.

Relation between the length scale (distance between non-decimated sites) and the energy-scale reads as:
\begin{equation}
\Omega \sim L^{-\frac{1}{2 |\Delta|}}
\;,
\label{lomega1}
\end{equation}
thus $\Delta$ is simply related to the dynamical exponent as $z=1/2 |\Delta|$.

The properties of the Griffiths phases can be interpreted in terms of
rare regions, as reviewed in \cite{vojta06,vojta13}.

\subsubsection{Consequences for the ground-state wavefunction}

The SDRG procedure
allows to evaluate any observable in each given disordered sample.
Besides all the observables of interest reviewed in \cite{igloi05review}, 
a particular attention has been paid in recent years 
towards the characterization of the ground-state wavefunction,
via its fidelity \cite{garnerone09} and its multifractality \cite{monthus15entropygs},
while the entanglement properties will be discussed in detail in Section \ref{sec:entanglement}.

\subsection{Higher dimensions $d>1$ }
\begin{figure}[h]
  \begin{center}
    \includegraphics[width=9cm]{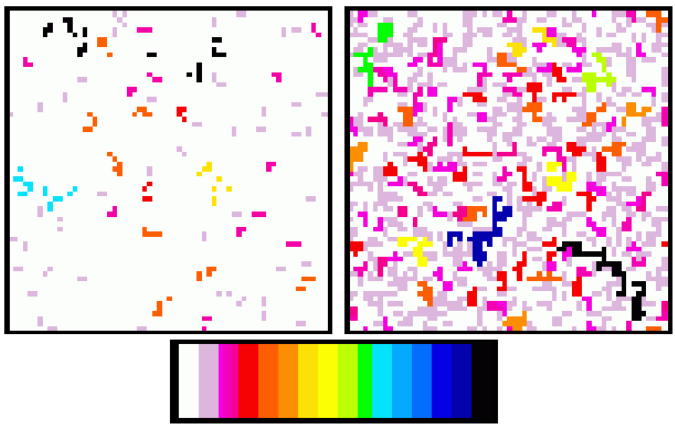}
  \end{center}
  \caption{Structure of connected clusters at the critical point of the 2D RTIM with $L=64$. Left panel: fixed-$h$ distribution, right panel: box-distribution. The colour-code is used to indicate the moment of the clusters.}
  \label{Fig1}
\end{figure}
In more complicated geometries, in particular in higher dimensions,
 the SDRG can in principle be performed numerically and the first results are reviewed in \cite{igloi05review}.
 In practical applications, a large finite sample with $N$ sites is renormalized up to the last effective spin,
 and the original sites of the sample are parts of effective clusters of different sizes. In the \textit{paramagnetic phase}, the clusters have a finite typical linear extent $\xi$, which characterizes the correlation length of the system. 
As the critical point is approached, $\xi$ diverges as $\xi \sim |\theta-\theta_c|^{-\nu}$. 
In the \textit{ferromagnetic phase} $\theta<\theta_c$, there is a huge connected cluster, which is compact and contains a finite fraction $m$ of the sites, that directly represents the average longitudinal magnetization. 
At the \textit{critical point}, the giant cluster is a fractal : its total moment $\mu$ scales with the linear size $L$ of the system
 as $\mu \sim L^{d_f}$, $d_f$ being the fractal dimension as in Eq.(\ref{d_f}). This is related to the scaling dimension of the longitudinal magnetization as $x=d-d_f$. Renormalizing sites at a special position (surfaces, edges, wedges and cones),
 one obtains the scaling behaviour of the system at the special local environment, which is characterised with the actual local scaling exponent\cite{kovacs13}. One can also consider more special geometries, such as multiple-junctions, at which several semi-infinite chains or plans meet \cite{juhasz14star,monthus15star}.

Transverse-spin correlations are defined as: $\langle \sigma_i^z \sigma_j^z \rangle$, which are of ${\cal O}(1)$, if both sites $i$ and $j$ are one-site clusters, (i.e. both are immediately decimated out, and these are the white sites in Fig.\ref{Fig1}) otherwise the correlation is negligible. At the critical point one should consider the connected part of the average transverse-spin correlation function, which decays with the distance as a power with the exponent $\eta_t$. Numerical SDRG results indicate\cite{igloi18}, that this exponent is approximately $\eta_t \approx 2+2d$, for $1 \le d \le 3$, see in Table~\ref{table:1}.

One important point is the relation between the energy-scale $\Omega$ (or inverse time-scale) and the length-scale: in the numerical procedure, $\Omega$ is the transverse field associated with the last effective spin. If the scaling is controlled by an IDFP, then the scaling relation is activated $\ln \Omega \sim L^{-\psi}$ as in Eq.(\ref{Psi}). For strong disorder Fixed Point, the scaling relation is in the conventional power-law form $\Omega \sim L^{-z}$, which generally holds in the Griffiths phase, and the dynamical exponent $z$ depends on the value of the control parameter, see also in Eq.(\ref{lomega1}). Distribution of the smallest gap in different random samples, $P(\Omega,L)$ depends on the variable $u=\Omega L^z$ :  it is universal and given by the limit distribution of extremes of independent and identically distributed random numbers\cite{galambos78}, as explained in the framework of the SDRG approach in Ref. \cite{juhasz06}.

\begin{table}[h]
\caption{Critical exponents of the RTIM in different dimensions. In 1D the analytical
results are from\cite{fisher92}, in 2D the numerical results are taken from\cite{kovacs10}, in 3D and 4D these are from\cite{kovacs11a,kovacs11b}.
The surface magnetization exponent $x_s$ and the decay exponent of the transverse magnetization, $\eta_t$ are from \cite{kovacs13} and \cite{igloi18}, respectively.\label{table:1}}
 \begin{tabular}{|c|c|c|c|c|}  \hline
   & 1D & 2D & 3D & 4D\\ \hline
$\nu$   &  $2.$ & $1.24(2)$ & $0.98(5)$ & $0.79(5)$  \\ 
$x$   &  $\frac{3-\sqrt{5}}{4}$ & $0.982(15)$ & $1.840(15)$ & $2.72(12)$  \\
$x_s$   &  $0.5$ & $1.60(2)$ & $2.65(15)$ & $3.7(1)$  \\
$\eta_t$   &  $4.1(2)$ & $6.0(2)$ & $8.1(2)$ &   \\
$\psi$   &  $1/2$ & $0.48(2)$ & $0.46(2)$ & $0.46(2)$  \\ \hline
  \end{tabular}
  \end{table}

The numerical implementation of the SDRG procedure needs some care. Naive application of the decimation rules leads to a computational time which scales as $t \sim {\cal O}(N^3)$. Such type of procedure has been used for ladders \cite{kovacs09} and for square samples of linear size up to $L \approx 160$ \cite{lin07,yu08}. In Ref. \cite{laumann12},
 the so called planar approximation is introduced and in this way they could go up to $L=500$. Finally, using the maximum rule, an efficient numerical procedure has been introduced in Refs. \cite{kovacs11a,kovacs11b} which works as $t \sim {\cal O}(N\ln N +E)$, where $E$ stands for the number of edges of the lattice. In models with nearest-neighbour interaction $E \sim N$ and one can go up to $N=4 \times 10^6$, c.f. in 2D up to $L=2048$. Numerical studies of the RTIM in 2D, 3D and 4D indicate, that in each case the critical behaviour is governed by an IDFP, like in the one-dimensional case. The probably most accurate values of the critical exponents are collected in Table \ref{table:1}. The critical exponents of the RTIM in any studied dimension have found universal, i.e. they do not depend on the actual form of the initial disorder.

To decide about the upper critical dimension of the IDFP, the critical properties of the RTIM have been studied on Erd\H os-R\'enyi random graphs \cite{kovacs11a,kovacs11b}, which are formally infinite-dimensional objects. The numerical results have indicated that the critical behaviour is controlled by a logarithmically infinite disorder fixed point, 
pointing towards an infinite upper critical dimension for this type of problem.

\subsection{Approximate RG methods}

Simpler approximation methods have been developed and applied to the
RTIM \cite{dimitrova11,monthus_garel1,monthus_surface,monthus_garel2,monthus_garel3,monthus_garel4,miyazaki12}.
One of those \cite{dimitrova11} is based on the quantum cavity approach \cite{ioffe10},
which is found to reproduce some of the exact results in 1D.
The quantum cavity method is equivalent to a linearized transfer matrix approach \cite{monthus_surface}.
 If no linearization is performed, i.e. in the so called non-linear transfer matrix approach,
one obtains IDFP behavior for $d \ge 2$ \cite{monthus_surface}.
Also approximate
renormalization group schemes have been suggested \cite{monthus_garel2,monthus_garel3,monthus_garel4,miyazaki12},
during which the order of the RG steps is changed in such a way
that the proliferation of new couplings is avoided. 
These methods have reproduced some exact 1D results and also provide IDFP behavior
for $d \ge 2$, in agreement with the standard SDRG method.

\section{Random quantum systems with long-range interactions}

\label{sec:LR}

In nature there are magnetic materials in which ordering is due to long-range (LR) interactions which decay as a power $\alpha=d+\sigma$ with the distance. The best known examples are dipolar systems, such as the $\mathrm{LiHoF}_4$. Putting this compound into an appropriate external magnetic field we obtain an experimental realisation of a dipolar quantum ferromagnet \cite{dutta15}. Similar systems have been experimentally realised recently by ultracold atomic gases in optical lattices \cite{friedenauer08,kim10,islam11,britton12,islam13}. Here we consider quantum magnets with LR interactions in the presence of quenched disorder. Such type of a system is realised
by the compound $\mathrm{LiHo}_x\mathrm{Y}_{1-x}\mathrm{F}_4$, in which a fraction of $(1-x)$ of the magnetic $\mathrm{Ho}$ atoms is replaced by non-magnetic $\mathrm{Y}$ atoms \cite{dutta15}. A related, but somewhat simplified quantum model which describes the low-energy properties of this system is the random transverse-field Ising model with LR interactions given by the Hamiltonian:
\be
{\cal H} =
-\sum_{i\neq j} \frac{b_{ij}}{r_{ij}^{\alpha}} \sigma_i^x \sigma_{j}^x-\sum_{i} h_i \sigma_i^z\;.
\label{eq:HLR}
\ee
where the $b_{ij}>0$ parameters and the $h_i>0$ transverse fields are i.i.d. random variables with given initial distributions.

In the LR model couplings and transverse fields play a different role and this asymmetry is manifested in the SDRG trajectories. At the critical trajectory couplings are very rarely decimated and the renormalised transverse fields follow Eq.(\ref{h_tilde}). On the contrary almost always transverse fields are decimated out, and according to the maximum rule in Eq.(\ref{J_tilde}) we have typically: $\tilde{J}_{jk}\approx J_{jk}$. Using these observations a primary model has been formulated, which has an exact solution in one dimension. 

\subsection{1D - solution of the primary model \cite{juhasz14a}}
\label{sec:primary}

In the primary model, the transverse fields are random,
but the couplings are non-random, i.e. $b_{ij}=b=1$. In the paramagnetic phase and at the critical point,
almost always transverse fields are decimated. After decimating $h_i$, the effective coupling between nearest clusters
$i-1$ and $i+1$ will always be smaller than the deleted ones, $J_{i-1,i}$ and 
 $J_{i,i+1}$ and we assume that according to the maximum rule $\tilde{J}_{i-1,i+1}={J}_{i-1,i+1}$.
Then the renormalization rule of couplings
between nearest clusters can be expressed in terms of the length variables as 
 $\tilde{J}_{i-1,i+1}^{-1/\alpha}=J_{i-1,i}^{-1/\alpha}+J_{i,i+1}^{-1/\alpha}
+w_i$, where $w_i$ is the extension of the cluster, which will be neglected in the following.
Using reduced variables 
$\zeta=\left(\frac{\Omega}{J}\right)^{1/\alpha}-1$ and
$\beta=\frac{1}{\alpha}\ln\frac{\Omega}{h}$, the approximate renormalization
rules are 
\be
\tilde\zeta=\zeta_{i-1,i}+\zeta_{i,i+1}+1\;
\label{zeta}
\ee
and
\be
\tilde\beta=\beta_i+\beta_{i+1}\;
\label{beta}
\ee
for field and bond decimation, respectively.
Since, in the ferromagnetic phase, the effective couplings between remote
clusters may be stronger than those between adjacent ones due to the large mass
of clusters, this approach is justified only in the paramagnetic phase and at the
critical point.

The decimation equations in Eqs.(\ref{zeta}) and (\ref{beta}) are identical to those of the $1d$ disordered $O(2)$ quantum rotor model of
granular superconductors \cite{altman04,altman10} with the grain charging energy 
$U_i$ and Josephson coupling $\mathcal{J}_{i,i+1}$  corresponding to 
$U_i\leftrightarrow J_{i,i+1}^{1/\alpha}$ and
$\mathcal{J}_{i,i+1}\leftrightarrow h_{i}^{1/\alpha}$, see in Sec.\ref{sec:superfluid-insulator}.

The fixed-point solution for the distributions $\zeta$ and $\beta$ can be found in \cite{juhasz14a}, here we recapitulate the basic results. The solutions are parameterised with a variable $a$, which is positive $a>0$ in the paramagnetic phase and 
vanishes $a=0$ at the critical point. The relation between the average distance of clusters, $L$, and the energy-scale $\Omega$ is given by:
\be 
L \sim
\left(\frac{\Omega_0}{\Omega}\right)^{\frac{1+a}{\alpha}}\;,
\label{dyn}
\ee
with an additional factor $\ln^2(\Omega/\Omega_0)$ for
$a=0$.
Thus the dynamical exponent $z=\alpha/(1+a)$ is a continuous function of $a$,
and it is maximal but finite at the critical point:
$z_c=\alpha$. 
The limit distribution of the transverse fields for $\Gamma \to \infty$ follows
the power law $g(h) \sim h^{1/z-1}$, thus the transition is controlled by a strong disorder fixed point.
The average correlation length scales at the vicinity of the critical point as:
\be
\xi \sim\exp(C'/a),\quad a=(\theta-\theta_c)\;,
\label{xi_KT}
\ee
which is similar to that at a Kosterlitz-Thouless transition point. The ratio of decimated couplings and decimated fields scales at the critical point with the size of the system, $L$, as:
\be
r(L)\simeq 2 \ln^{-\omega} (L/L_0), \quad \omega=2\;.
\label{r(L)}
\ee
The typical magnetic moment of critical clusters scales also logarithmically:
\be
\mu(L) \sim \ln^{\chi}L, \quad \chi=2\;,
\label{mu(L)}
\ee
thus the fractal dimension is formally $d_f=0$. Finally, the entanglement entropy $S_L$ of a finite block of size
$L$ in an infinite system is found to approach a finite limiting value as $L \to \infty$, even at the critical point.

Results of the numerical SDRG analysis (with the maximum rule) of the LR random transverse-field Ising chain are in agreement with the findings of the primary model, even if the couplings, i.e. the parameters $b_{ij}$ are random.

\subsection{3D - numerical Strong Disorder RG study}

The 3D LR model has been studied by the SDRG approach with the maximum rule \cite{kovacs16} and similar behaviour of the RG trajectories are observed as in 1D. At a given energy scale, $\Omega$, the renormalization is characterised by the ratio of decimated couplings and decimated fields, $r$. 

In the paramagnetic phase and at the critical point, where the maximum rule is expected to hold, at the line of fixed points, $\Omega \to 0$ we have $r \to 0$, and these fixed points are stable. In the vicinity of this fixed line almost exclusively transverse fields are decimated, the distribution of which is given by a power-law:
\begin{equation}
g(h) = \frac{d}{z}h^{-1+d/z}\;,
\label{h_distr}
\end{equation}
with an effective ($\Omega$, i.e. $r$ dependent) dynamical exponent $z$. At the fixed line $z$ is maximal at the critical point, having a value $z_c \approx \alpha$, as in 1D. The other critical parameters (correlation length, decimation ratio
and cluster moment) have similar behaviour as in 1D, thus the relations in Eqs.(\ref{xi_KT},\ref{r(L)},\ref{mu(L)}) are valid, only the exponents of the logarithm in Eqs.(\ref{r(L)}) and (\ref{mu(L)}) are somewhat different.

\begin{figure}[th]
\begin{center}
\includegraphics[width=6.5cm,angle=0]{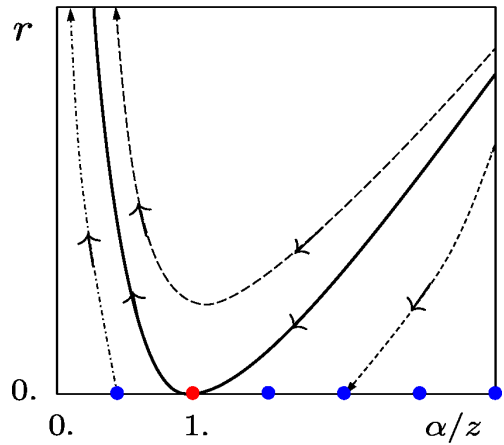}  
\end{center}
\vskip -.5cm
\caption{\label{fig_2} (Color online) Schematic SDRG phase diagram obtained through the maximum rule as a function of the ratio $r$ and the effective dynamical exponent defined in Eq.(\ref{h_distr}). The arrows indicate the direction the parameters evolve as the energy-scale is reduced. Fixed points (blue circles) are at $r=0$: the attractive fixed points of the paramagnetic phase ($\alpha/z>1$) and the repulsive ones ($\alpha/z<1$) are separated by the critical fixed point (red circle).}
\end{figure}

The RG phase-diagram can be extended to include the ferromagnetic phase, too, where the RG-flow scales to $r \to \infty$. This is shown in Fig.\ref{fig_2}. The line of fixed points at $r=0$ at other side of the critical point with $\alpha/z<1$ are unstable and the RG-flow scales to $r \to \infty$. In this regime the maximum rule in the SDRG procedure is certainly not valid. The two regimes of fixed points are separated by the critical fixed point at $\alpha/z=1$.

In the vicinity of the line of fixed points at $r=0$, almost exclusively transverse fields are decimated out. In a finite cluster of linear size  $\ell$ however, there are the few smallest ones which remain intact. From the distribution of the fields in Eq.(\ref{h_distr}) one can estimate the value of these non-decimated fields through extreme value statistics. This type of analysis leads to basically identical results, as obtained through the SDRG approach. Using this type of analysis it has been argued that the extrapolated value of the magnetization from the ferromagnetic side has a finite limiting value and thus the transition is of mixed-order.

\subsection{Other quantum models with LR interactions}

\subsubsection{Ising chains with other type of LR interactions}

The case where the Long-ranged interactions are dilute are analyzed in \cite{dutta06LRdilute}.
The Dyson hierarchical version of the quantum Ising chain with LR power-law ferromagnetic couplings
 and pure or random transverse fields is studied via real-space renormalization in \cite{monthus15b}, 
while the Dyson Hierarchical LR  Quantum Spin-Glasses are investigated in \cite{monthus15c}.

The critical properties of random quantum systems, in particular the RTIM in one dimension in the presence of LR 
interactions which decay in a stretched exponential fashion, as $J(r) \sim \exp(-C r^{a})$ has been studied in Ref. \cite{juhasz14b}. Using a variant of the SDRG approach similar to the primary model presented in Sec.\ref{sec:primary}, the critical behaviour is found to depend on the parameter $a$. For $0<a<1/2$ the critical behaviour is controlled by an IDFP, in which the critical exponents are $a$-dependent and these are different from those in the SR model. For example one obtains $\psi=a$, which is understandable, since the relation between energy-scale and length-scale is dominated here the form of LR interactions. On the contrary for $a>1/2$ the LR interactions are irrelevant and the critical properties of the model are the same as its SR variant.

\subsubsection{Random Heisenberg chains with LR interactions}

In Refs. \cite{moure15,moure18} the Hamiltonian of $N$ LR interacting $S=1/2$ spins is considered:
\be
{\cal H}_{Heis}=\sum_{i \ne j, \beta} J_{ij}^{\beta} S_i^{\beta}S_j^{\beta}\;,
\ee
which are placed randomly on a periodic lattice of length $L$ and lattice spacing $s$. The interaction between all pairs of sites $i,j$ are antiferromagnetic:
\be
J_{ij}^{\beta}=J|(r_i-r_j)/a|^{-\alpha} \exp(-|(r_i-r_j)|/\xi)\;,
\ee
having a power-law decay with exponent $\alpha$ and an exponential cut off with a length scale $\xi$. This model is aimed to better understand the magnetic properties of doped semiconductors. The singular properties of the model are studied by the SDRG method with the use of the sum rule. In the LR limit, $\xi \to \infty$, the distribution function of the low energy excitations are studied by various values of $\alpha$ and at a critical value $\alpha_c=1.6$ it coincides with a critical function pointing towards the existence of a many body localization transition. 

In Ref. \cite{moure18} the magnetic susceptibility $\chi(T)$ is studied as a function of the temperature $T$. In the LR limit, $\xi \to \infty$ a crossover is observed at $\alpha^*=1.066$ between a phase with a divergent low-temperature susceptibility $\chi(T \to 0)$ for $\alpha > \alpha^*$ to a phase with a vanishing $\chi(T \to 0)$ for $\alpha < \alpha^*$.


\section{ Other random quantum models }

\label{sec:other}

The RTIM is one of the most studied random quantum models, for wich some new quantities, such as the transverse magnetization\cite{igloi18} and the magnetic Gr\"uneisen ratio\cite{vojta10gruneisen} have been studied by the SDRG method.
Besides models that can be directly mapped onto the RTIM,
such as for instance the Majorana model considered in \cite{shivamoggi10}, many other short-ranged 
random quantum models
have been analyzed via SDRG since the Review \cite{igloi05review}.

\subsection{Models with discrete symmetry}

\subsubsection{Antiferromagnetic random quantum Ising chain}

The antiferromagnetic random quantum Ising chain is defined by the Hamiltonian:
\be
{\cal H}_{AF}=\sum_i J_i \sigma_i^z \sigma_{i+1}^z - \sum_i h_i \sigma_i^x - \sum_i H_i \sigma_i^z\;,
\ee
with $J_i,h_i,H_i > 0$. The clean model has paramagnetic and antiferromagnetic ordered phases, which are separated by a critical line, which for $h>0$ is controlled by the Ising quantum fixed point at $h/J=1$ and $H/J=0$. At $h=0$, when the model is classical there is a multicritical point at $H/J=2$ separating the antiferromagnetic phase from the ferromagnetic one.

In the random chain, where the couplings are distributed uniformly in $0<J<1$ and the random transverse-fields 
are distributed uniformly in $0<h<h_0$ (but the longitudinal fields were non-random, $H_i=H$), infinite disorder scaling is observed only at $H=0$, which is equivalent to the RTIM in 1D. For any finite value of $H>0$ the numerical results indicate strong disorder scaling, thus a paramagnetic phase with Griffiths singularities \cite{lin17}. More recent studies indicate, that the region of infinite disorder criticality is extended to $H>0$, provided the distributions of the couplings and that of the transverse fields have a finite bounding value from below and above \cite{lajko18}.

\subsubsection{Ashkin-Teller chains}

The $N$-colour Ashkin-Teller chain is defined in terms of $\alpha=1,2,\dots,N$-sets Pauli-matrices by the Hamiltonian:
\beqn
&{\cal H}_{AT}=-\sum_{\alpha=1}^N\sum_i \left(J_i \sigma_{\alpha,i}^z \sigma_{\alpha,i+1}^z + h_i \sigma_{\alpha,i}^x \right) \nonumber \\
&-\sum_{\alpha<\beta}^N\sum_i \left(K_i \sigma_{\alpha,i}^z \sigma_{\alpha,i+1}^z \sigma_{\beta,i}^z \sigma_{\beta,i+1}^z+g_i \sigma_{\alpha,i}^x \sigma_{\beta,i}^x\right)\;,
\eeqn
which is the generalization of the standard Ashkin-Teller chain with $N=2$, which has been studied before and reviewed in  \cite{igloi05review}. In terms of positive random parameters, however keeping the ratios $\epsilon_{h,i}=g_i/h_i$ and $\epsilon_{J,i}=K_i/J_i$
spatially homogeneous: $\epsilon_{h,i}=\epsilon_{J,i}=\epsilon_i$ the model has been investigated by the SDRG method in a series of papers \cite{goswami08,hrahsheh12,hrahsheh14,barghathi15} Different phases and various critical and multicritical points have been identified, which all are of the infinite disorder type. Various numerical studies 
have been performed to confirm the SDRG predictions \cite{zhu15,bellafard16,chatelain16,ibrahim17}.

\subsection{Models with continuous symmetry}

The random Heisenberg Antiferromagnetic spin chain is the first model where SDRG has been introduced \cite{ma79}.
After the various works already reviewed in \cite{igloi05review}, more recent studies include
the effects of next-nearest-neighbor interaction in $d=1$ \cite{lamas06},
the case of wealkly coupled chains \cite{kokalj},
models in dimension $d=2$ \cite{yu06AF2d,ma14AF2d,liu18AFd2},
as well as the generalizations to various type continuous symmetry ($SU(3)$, $SU(N)$, $SO(N)$) 
considered in the series of papers\cite{zhou09,quito15,quito16b,quito17a,quito17b},
 where different types of random singlet phases are identified via SDRG and the low-energy behaviour is controlled by infinite disorder fixed points. 

The $S=1$ random spin chain is studied by the numerical application of the SDRG method \cite{lajko06},
finding that 
the cross-over in the critical behaviour with the strength of disorder is in agreement with the analytical theory .
Other studies of random spin chains with string-order parameter
include the random Majumdar-Ghosh Chain \cite{lavarelo12}
and the random Cluster-Ising-Model \cite{strinati17}.

The one-dimensional Hubbard model with random-hopping matrix-elements, and with random onsite Coulomb repulsion terms is studied by the SDRG method \cite{melin06}.
 Two critical phases are identified, which correspond to an infinite disorder
spin random singlet for strong interactions and to an orbital infinite disorder fixed point for vanishing interactions. To each critical infinite disorder fixed point is connected a Griffiths phase.

Layers and bilayers of S=1/2 Heisenberg antiferromagnets with different types of disorder: bond randomness, site dilution, and dimer dilution are studied by the numerical application of the SDRG method \cite{lin06}. Generally the systems exhibit an ordered and a disordered phase separated by a phase boundary on which the static critical exponents appear to be independent of bond randomness in the strong-disorder regime, while the dynamical exponent is a continuous function of the bond disorder strength. The low-energy fixed points of the off-critical phases are affected by the actual form of the disorder, and the disorder-induced dynamical exponent depends on the disorder strength. As the strength of the bond disorder is increased, there is a set of crossovers in the properties of the low-energy singularities.

\subsection{Disordered non-abelian anyonic chains}

Non-abelian order is a particular feature in two-dimensional quantum systems and non-abelian excitations are present in fractional quantum Hall states. Chains of interacting anyonic quasiparticles are introduced recently and their properties in the presence of quenched disorder has been studied through the SDRG method \cite{bonesteel07,fidkowski08a,fidkowski08b,refael09}.

Let us consider a simple example of anyonic models, the Fibonacci or golden chain, which is introduced on the analogy of the $S=1/2$ Heisenberg model with $SU(2)$ symmetry: ${\cal H}=\sum_i J_i {\mathbf S}_i {\mathbf S}_{i+1}$, for which the energy of a nearest neighbour pair is different in the singlet and in the triplet chanels, and according to tensor product or the fusion rule we have: $\frac{1}{2} \otimes \frac{1}{2}=0 \oplus 1$. The Fibonacci chain consists of non-abelian anyons carrying a topological charge, $\tau$, and their interaction is described by the fusion rule: $\tau \otimes \tau=1 \oplus \tau$, which means that the Hilbert space of two neighbouring anyons is the direct sum of unity and a copy of $\tau$. This type of construction can only be described by a truncated tensor product, where the $SU(2)$ representations are truncated at a level $k$. The Hamiltonian of the Fibonacci chain is given in the form:
\be
{\cal H}=\sum_{i=1}^N J_i P_i^A
\ee
where $J_i$ are the random interactions and $P_i^A$ is the singlet projection operator between site $i$ and $i+1$. Generally the Hilbert space of $N+1$ anyons is given by the $N^{\rm th}$ Fibonacci number.
Other type of non-abelian anyonic chains can be constructed in similar way, important class being the $SU(2)_k$ anyonic chains, the Fibonacci chain corresponds to $k=3$.

In the SDRG procedure we choose the strongest coupling in the chain, $\Omega=|J_i|$, and decimate it. For a strong antiferromagnetic coupling the sites $i$ and $i+1$ form a singlet, and an effective coupling $\tilde{J}\approx \kappa \frac{J_{i-1}J_{i+1}}{J_i}$ is formed between sites $i-1$ and $i+2$, with $\kappa=\phi^2$ and $\phi=\frac{1+\sqrt{5}}{2}$, the golden mean ratio. For a strong ferromagnetic bond sites $i$ and $i+1$ form a cluster, having effective couplings to the nearby sites: $\tilde{J}\approx-\frac{J_{i \pm 1}}{\phi}$.

The SDRG transformation has an infinite disorder fixed point, so that the ground state is a random singlet phase. The relation between energy-scale and length-scale is activated, with the critical exponent $\psi_{AF}=1/2$
if the original model is antiferromagnetic,
 and $\psi_{AF/F}=1/3$ if in the original model there is a finite fraction of ferromagnetic bonds. 
The SDRG analysis has been extended for other non-abelian anyonic chains, a detailed analysis of the results can be found in Ref. \cite{refael09}. Relation with $SU(N)$ symmetric random chains has been noticed in \cite{zhou09,quito15,quito16b,quito17a,quito17b}.

\subsection{Superfluid-insulator transition}
\label{sec:superfluid-insulator}

One-dimensional Josephson junction array with random couplings $J_i$ and random charging energies $U_i$ are described by a quantum rotor Hamiltonian:
\be
{\cal H}=\sum_i U_i(\hat{n}_i-\overline{n}_i)^2-\sum_i J_i \cos(\hat{\theta}_i-\hat{\theta}_{i+1})\;,
\ee
with charges $\hat{n}_i$ and phases $\hat{\theta}_i$ at site $i$, which satisfy the commutation relations: $\left[\hat{n}_k, \hat{\theta}_l\right]=-i \delta_{kl}$. At each site there is a random offset charge (or chemical potential), which is taken as $-1/2< \overline{n}_i <1/2$ and the integer part is absorbed into the definition of $\hat{n}_i$. This model is considered as an effective theory of interacting bosons propagating in a random potential and the type of the insulating phase is found to depend on the symmetry properties of the offset charge distribution.

The SDRG treatment of this model is described in Refs.  \cite{altman07,gurarie08,altman09,vosk12superfluid,iyer12,iyer13} and has been recently reviewed in\cite{refael13}. Here we just recapitulate the main ideas. The energy gap due to charging energy is $\Delta_i=U_i(1-2|\overline{n}_i)$, that should be compared with the interaction coupling $J_i$, and their maximum term defines the energy-scale $\Omega$.

If $\Omega=\Delta_i$ corresponds to a large gap, this site is fixed to its lowest energy state and an effective coupling is generated between sites $i-1$ and $i+1$:
\be
\tilde{J}=\frac{J_{i-1} J_i}{\Omega(1+2|\overline{n}_i|)}\;,
\ee
and in the denominator the term $1+2|\overline{n}_i|$ is set to unity. On the contrary,
if $\Omega=J_i$ corresponds to a large Josephson coupling, a cluster with a coherent phase is formed with an effective capacitance: $\tilde{C}=C_i+C_{i+1}$, thus the effective charging energy follows the rule:
\be
\frac{1}{\tilde{U}}=\frac{1}{U_i}+\frac{1}{U_{i+1}}\;.
\ee
Introducing scaling variables: $\beta_i=\ln(\Omega/J_i)$ and $\xi_i=\Omega/U_i-1$ in the vicinity of the fixed point, where the expressions for the offset charges are simplified, one arrives at the same form of the flow equation of the distribution functions as for the primary model of the one-dimensional RTIM with LR interactions of Sec.\ref{sec:primary}. The paramagnetic phase of the LR RTIM corresponds to the superfluid phase, while the ferromagnetic one to the insulator phase. The superfluid phase is just a Griffiths phase with a dynamical exponent $z=1/(1+a)$, the supefluid-insulator transition is of strong disorder type with $z=1$ and with an exponentially diverging correlation length. Details of the solution and a review of the numerical results are given in Ref. \cite{refael13}.
As a final remark, let us stress that the nature of Superfluid-Bose Glass Transition has remained controversial over the years,
as summarized in recent numerical investigations\cite{hrahsheh12,doggen17}.

\subsection{Superconductor-Metal transition}

An Infinite-Randomness Fixed Point has also been found for the
Superconductor-Metal Quantum Phase Transition \cite{maestro08,maestro10,nozadze14,ibrahim18}.

\subsection{The Rainbow Spin Chain}

Although the so-called 'Rainbow spin chain' is not random, the spatial structure of its inhomogeneity
allows to apply iteratively the Ma-Dasgupta SDRG rule to construct the ground state and analyze its entanglement properties
\cite{ramirez14,ramirez15,rodriguez16,rodriguez17,ramirez18}.
The addition of disorder in this rainbow spin chain has been also studied recently via SDRG \cite{alba18}.

\section{ Entanglement properties }

\label{sec:entanglement}

The entanglement of quantum many body systems is a promising concept to understand
their topological and universal properties, in particular in the vicinity of a quantum phase-transition point (see the reviews \cite{calabrese09,amico08,laflorencie16}).
The entanglement of the ground state $\left|\varPsi\right\rangle$
between the subsystem ${\cal A}$ and the rest ${\cal B}$ of the system
is quantified by the von Neumann entropy ${\cal S}=-{\rm Tr}_{\cal A}\left(\rho_{\cal A} \log_2{ \rho_{\cal A}}\right)$
 of the reduced density matrix 
$\rho_{\cal A}={\rm Tr}_{{\cal B}} | \varPsi \rangle \langle \varPsi |$.
Generally ${\cal S}$ scales with the area of the interface separating ${\cal A}$ and ${\cal B}$.
In some cases however, there are singular corrections to the area law.
In one-dimensional pure systems, ${\cal S}$ is logarithmically
divergent at a quantum critical point\cite{holzhey94,vidal03,calabrese04}: ${\cal S}=\frac{c}{3} \log_2 \ell+cst$.
 Here $\ell$ is the size of the subsystem ${\cal A}$
and the prefactor is universal, $c$ being the central charge of the conformal field theory. 
Besides this entanglement entropy, it is interesting to consider also the full entanglement spectrum \cite{calabrese08}.

In random quantum systems the entanglement properties are conveniently studied by the SDRG approach and many of the obtained results, mainly in one-dimension are thoroughly reviewed in Refs. \cite{refael09,laflorencie16}. Therefore here we just shortly mention the known results in 1D and we concentrate on the higher dimensional results, as well as on the more recent developments.

\subsection{Random quantum chains}

The ground state of the $S=1/2$ random-bond Heisenberg chain in the SDRG approach is a random singlet phase and each singlet bond which connects the two subsystems ${\cal A}$ and ${\cal B}$ has a contribution $1$ to the entanglement entropy. The entanglement entropy scales logarithmically: ${\cal S}=\frac{c_{eff}}{3} \log_2 \ell+cst$, with an effective central charge $c_{eff}=\ln 2$. The entanglement across a weakened link is studied in Ref. \cite{vasseur17}, while 
the multifractal Orthogonality Catastrophe produced by a local cut is analyzed in \cite{vasseur15multif}.

The ground state of random $S>1/2$ chains depends on the strength of disorder. For strong enough disorder it is a spin-$S$ random singlet phase having an effective central charge $c_{eff}=\ln(2S+1)$\cite{hoyos07,refael07,saguia07,saguia10}. For weaker disorder there are a set of multi-critical points, at which the central charge has not yet been calculated exactly. The case $S=1$ is studied in  Ref. \cite{refael07}.

For the RTIM the ground state in the SDRG approach consists of a set of clusters of different sizes (see in Fig.\ref{Fig1}) and each cluster which has points in both subsystems ${\cal A}$ and ${\cal B}$ has a contribution $1$ to the entanglement entropy. In one dimension, according to SDRG calculations in the off-critical region,
 ${\cal S}$ is finite and thus satisfies the area law. At the critical point, it is logarithmically divergent, and the effective central charge is $c_{eff}=\ln 2/2$, i.e. just the half of that in the random XX- and S=1/2 Heisenberg chains, which follows also from an exact mapping \cite{igloi08}. In a given sample of finite length, the position of the maximum of the average entanglement entropy (the average is made over all possible positions of the subsystem) 
can be used to define a sample dependent pseudo-critical point \cite{igloi07}.

The effective central charge of random one-dimensional systems calculated from the average entanglement entropy is generally smaller than its analogous value in the pure model. One can however construct models, in which the effective central charge of the random model is the larger \cite{santachiara06}. One can also construct models with (locally) correlated disorder, so that the local control parameter stays constant \cite{hoyos11}. In such models the critical behaviour, as well as scaling of the critical entanglement entropy follows the same form as in the pure systems \cite{hoyos11,getelina16}.

The entanglement entropy between a random and a clean part of a system, such in the XX-chain has also been considered recently. In this case, at the critical point, a very weak, double-logarithmic dependence is observed \cite{juhasz17a}: ${\cal S} \sim \ln \ln \ell$. For a related study of entanglement across extended random defects see \cite{juhasz17b}. Entanglement entropy at multiple junctions of random quantum chains has been studied in \cite{juhasz18}.
The SDRG approach has also been used to study the critical properties of aperiodic quantum spin chains \cite{vieira05,filho12,casa14,vieira18} and their entanglement entropy is calculated in the strong aperiodicity limit \cite{igloi07b,juhasz07b}.

Other entanglement measures have been also studied in random quantum chains,
like the entanglement \cite{hoyos06channels,getelina17} or the concurrence \cite{hide11} between distant pairs of q-bits,
the full entanglement spectrum of random singlet critical points \cite{fagotti11}, 
the R\'enyi entropies \cite{ramirez14ent},
the fluctuations of the entanglement entropy \cite{tran09},
the full probability distribution of the entanglement entropy \cite{devakul17},
the Schmidt-gap ( i.e. the difference between the two largest eigenvalues of the entanglement spectrum) for the RTIM and for the $S=1$ random spin chain \cite{torlai18}. Using the SDRG method the entanglement negativity in random singlet phases are shown to scale logarithmically with the size of the system \cite{ruggiero16}.

Motivated by the entanglement entropy in the random singlet phase of the random $S=1/2$ spin chain, entanglement measure through valence bond entanglement has been proposed for $SU(2)$ quantum systems \cite{alet07}, which can be efficiently measured through quantum MC calculations both in one- and two-dimensions \cite{lin10,shu16}. 

\subsection{RTIM in higher dimensions}

The entanglement entropy of the RTIM is given by the number of such clusters in the ground state which have points both in ${\cal A}$ and ${\cal B}$. This quantity has been considered first in Ref. \cite{lin07} and at the critical point, a singular behavior in the form of ${\cal S} \sim \ell \ln \ln \ell$ has been found in finite systems with linear size $L=64$. Soon after the calculation has been extended up to $L=160$ \cite{yu08} and the numerical results have been interpreted as a logarithmic correction to the are law: ${\cal S} \approx a \ell +b \ln \ell$. To decide between the two suggested singular forms, a calculation has been performed with the very efficient numerical algorithm up to $L=2028$ and by two different forms of disorder. These calculations have been performed also for 3D and 4D \cite{kovacs12a}.

At the critical point of a D-dimensional system, the entanglement entropy when the subsystem ${\cal A}$ is a cube is found to be in the form:
\be
{\cal S}_{\rm cube}^{(D)}(\ell)=a_{D-1} f_{D-1} + \sum_{E=1}^{D-2} a_E f_{E} + {\cal S}_{\rm cr}^{(D)}(\ell)\;,
\label{S^d}
\ee
where the first term represents the area-law, the second terms are analytical corrections due to $E$-dimensional edges and the last term is the corner contribution, which is logarithmically divergent: ${\cal S}_{\rm cr}^{(D)}(\ell)=b^{(D)}\ln \ell + const$. According to numerical estimates the prefactors are universal, i.e. disorder independent and given by $b^{(2)}=-0.029(1)$, $b^{(3)}=0.012(2)$ and $b^{(4)}=-0.006(2)$.

The corner-entropy has also been studied in the vicinity of the critical point and found to be extremal at the critical point. Outside the critical point, ${\cal S}_{\rm cr}^{(D)}(\ell,\delta)$ is finite and can be obtained by replacing $\ell$ with the finite correlation length $\xi$, where the divergence of $\xi$ at the critical point is characterised by the given exponent in Table~\ref{table:1}.

\subsection{Bond diluted quantum Ising model}

The bond diluted quantum Ising model is defined by the Hamiltonian in Eq.(\ref{eq:H}) with $h_i=h$ and with nearest-neighbour couplings which are $J>0$ with probability $p$ and $J=0$ with probability $1-p$. At the percolation transition point $p_c$, for small transverse field $h$, there is a line of phase transition, the critical properties of
which are controlled by the percolation fixed point \cite{senthil96}, for a review see \cite{vojta07}. The ground state of ${\cal H}$
is given by a set of ordered clusters, which are in the same form as for percolation. Now consider
a subsystem ${\cal A}$ with boundary $\Gamma$ and calculate the entanglement
entropy between the subsystem and the environment, which is given by the number of clusters in ${\cal A}$ which intersect $\Gamma$ and contain also at least one point of the environment. 

In two dimensions it is given in the same form as for the RTIM \cite{kovacs12b}:
\be
 {\cal S}_{\Gamma}=a L_{\Gamma}+b\log L_{\Gamma} \;,
\label{S_Gamma}
\ee
where $L_{\Gamma}$ is the length of $\Gamma$. The prefactor of the logarithm in Eq.(\ref{S_Gamma}) is given by
the Cardy-Peschel formula \cite{cardy88}:
\beqn
b =&-&\dfrac{5 \sqrt{3}}{96 \pi} \sum_k \left[ \left( \dfrac{\pi}{\gamma_k}\right)- \left( \dfrac{\gamma_k}{\pi}\right)\right. \cr
&+&\left. \left(\dfrac{\pi}{2 \pi-\gamma_k}\right)-\left(\dfrac{2 \pi-\gamma_k}{\pi}\right)\right]\;,
\label{cardy_peschel}
\eeqn
where $\gamma_k$ is the interior angle at each corner. In the special case of the square subsystem one has $b=-5 \sqrt{3}/(36 \pi)=-0.07657$. The conformal prediction in Eq.(\ref{cardy_peschel}) has been confirmed by numerical calculations for different shapes of $\Gamma$ \cite{kovacs12b}.

In $d=3$, numerical calculations have confirmed that the singular contributions to the entanglement entropy of the bond diluted quantum Ising model are due to corners, and the prefactor of the logarithm is given by $b=1.72(3)$, which is different from that in the RTIM \cite{kovacs14}.

\subsection{ Relations between SDRG and Entanglement-Algorithms }

Since Tensor-Networks have become very popular in recent years,
it is interesting to point out that the SDRG actually corresponds to a special type of 
Multi-scale-Entanglement-Renormalization-Ansatz (MERA) (see section IV of the review \cite{cirac09})
and has been integrated into various tensor-network algorithms \cite{gittsovich10, goldsborough14,lin17,goldsborough17,chatelain18}.
Recently, in analogy with SDRG, a Strong-Disorder-Disentangling procedure \cite{hyatt17}
has been introduced : at each step, one chooses the most strongly entangled pair of sites,
 in order to construct iteratively the appropriate unitary circuit that transforms a given quantum state into an unentangled product state. The goal is to reveal the emergent entanglement geometry.

\section{ Localized and Many-Body-Localized Phases of quantum spin chains }

\label{sec:MBL}

For random quantum spin chains that can be mapped onto free-fermions via the Jordan-Wigner transformation,
the presence of disorder in this one-dimensional geometry leads to the Anderson-real-space-localization of all fermionic modes.
 In the presence of interactions, the issue of Many-Body-Localization in isolated random quantum spin chains
has attracted a lot of attention recently, as reviewed in \cite{nandkishore15,altman15mblreview,parameswaran17,alet18,abanin18}.
Among the various methods that have been proposed to construct the Local Integrals of Motion (LIOMs)
that characterize the Many-Body-Localized-Phase (see the reviews \cite{imbrie17,rademaker17} on LIOMs),
Strong Disorder RG procedures have been introduced under the names of RSRG-X and RSRG-t,
in order to construct the excited eigenstates or the effective dynamics respectively.

\subsection{ RSRG-X for excited eigenstates }
\label{sec:RSRG-X}

In order to construct the whole set of eigenstates,
the main idea of the RSRG-X procedure \cite{pekker14rsrgx}
is to keep the two possible local-energy-branches at each step,
instead of projectiong systematically onto the lowest local energy branch 
when the goal is to construct only the ground-state.
The RSRG-X can be formulated for the most general Hamiltonian involving Pauli matrices \cite{you16}
\begin{eqnarray}
H && = \sum_{[\mu]} h_{[\mu]} \sigma^{[\mu]} =  \sum_{\mu_1,..,\mu_N} h_{[\mu_1,..,\mu_N]} \sigma_1^{\mu_1} \sigma_2^{\mu_2} ...\sigma_N^{\mu_N} 
\label{hpauli}
\end{eqnarray}
where $\mu_i=0,1,2,3$ is the index of the Pauli matrix acting on spin $i$.
One chooses the maximum $\Omega= {\rm max} (h_{[\mu]} )= h{[\mu_0]} $ among the real couplings $h_{[\mu]}  $ of the Hamiltonian.
The corresponding term
\begin{eqnarray}
H_0 && =  h_{[\mu_0]} \sigma^{[\mu_0]} 
\label{mu0}
\end{eqnarray}
has two levels $(\pm h_{[\mu_0]} )$ corresponding to the high/low energy sectors.
The rest of the Hamiltonian can be classified according to the commutativity or anticommutativity with $H_0$
\begin{eqnarray}
H-H_0 && =  H_1^{comm}+H_1^{anti}
\label{commanti}
\end{eqnarray}
The part $H_1^{comm} $ that commutes with $H_0$ is kept to describe its effect withing each energy-level of $H_0$.
The part $H_1^{anti} $ that anticommutes with $H_0$ and that couples the two sectors is taken into account by second-order perturbation theory
to obtain the renormalized Hamiltonian within each energy sector of $H_0$
\begin{eqnarray}
H^R=H_0 +  H_1^{comm}-  H_1^{anti} \frac{1}{2 H_0}  H_1^{anti} 
\label{hrsrgx}
\end{eqnarray}
These rules are thus formally very similar to the Fisher SDRG rules for the ground state.
An alternative formulation of the RSRG-X rules in terms of Majorana fermions is described in \cite{monthus18rsrgxmaj} with its advantages.

While the writing of RSRG-X rules is a direct generalization of the SDRG-rules for the ground-state,
it should be stressed that their numerical implementation is much more involved.
Indeed, the exact construction of the $2^N$ eigenstates for a chain of $N$ spins is limited to small sizes
as a consequence of the exponential cost. To overcome this limitation, 
the authors of Ref \cite{pekker14rsrgx} have thus proposed to replace the exact application of the renormalization rules on all branches
by a Monte Carlo sampling of the typical branches of the tree.
The RSRG-X procedure has been applied to many random models,
including the XX chain \cite{huang14,pouranvari15},
the XXX chain \cite{agarwal15},
the XXZ chain \cite{vasseur16particlehole},
the XYZ chain \cite{slage16},
the three-state quantum clock model \cite{friedman17}
and anyonic spin chains \cite{vasseur15hot,kang17}.
Variants of the RSRG-X procedure have been also introduced to analyze the phase transition between different
MBL-phases in the Long-Ranged quantum spin-glass model \cite{monthus16emergent}
and for the random Transverse Field Spin-Glass Model on the Cayley tree \cite{monthus17mblcayley}.

\subsection{ RSRG-t for the unitary dynamics}
\label{sec:RSRG-t}

The RSRG-t for the effective unitary dynamics of isolated Quantum Spin chains \cite{vosk13,vosk14}
is based on the iterative elimination of the highest local frequency $\Omega$.
The idea is that the local degree of freedom with the two energy-levels $e_1$ and $e_2$ and corresponding projectors $P_{1,2}$ 
\begin{eqnarray}
H_0 = e_1 P_1+e_2 P_2
\label{h0twolevel}
\end{eqnarray}
that is associated to the highest frequency $\Omega=e_2-e_1$ oscillates freely.
In the interaction picture, the rest of the Hamiltonian
\begin{eqnarray}
V\equiv H-H_0
\label{vhhzero}
\end{eqnarray}
becomes the time-periodic Hamiltonian of high frequency $\Omega$
\begin{eqnarray}
V^{int} (t)  = e^{i H_0 t} V e^{-i H_0 t}  = {\cal V}_0 + {\cal V}_1 e^{i  \Omega t}+ {\cal V}_{-1} e^{-i  \Omega t}
\label{vhhtime}
\end{eqnarray}
with the three Fourier coefficients
\begin{eqnarray}
{\cal V}_{0}  && =    P_1 V   P_1+ P_2 V   P_2
\nonumber \\
{\cal V}_{1}  && =    P_2 V   P_1 
\nonumber \\
{\cal V}_{-1}  && = P_{1} V   P_{2}  =   {\cal V}_{1}^{\dagger} 
\label{vlnega}
\end{eqnarray}
The high-frequency-expansion for Floquet dynamics \cite{bukov15}
then yields that the effective Hamiltonian for the remaining degrees of freedom reads \cite{monthus17rsrgt}
\begin{eqnarray}
&& V_{eff} 
 =  \left(  P_1 V   P_1+ P_2 V   P_2 \right) 
\nonumber \\
&& + \frac{1}{\Omega} \left(   P_2 V   P_1    V   P_2  -     P_1 V   P_2 V   P_1 \right)
+ O\left(\frac{1}{\Omega^2} \right)
\label{vefffloquet}
\end{eqnarray}
The first ligne corresponds simply to the projection of $V$ onto the energy levels of $H_0$,
while the second ligne of order $1/\Omega$ contains virtual processes between the the two energy levels.
This formula for the RSRG-t rules is thus equivalent to the RSRG-X rules of Eq \ref{hrsrgx} based on 
the two first order perturbation theory for energy-levels, 
but this dynamical point of view shed a different light on the interpretation of the RG procedure.
At some given time $t$, the degrees of freedom are separated into two groups with respect to $\Omega_t = \frac{1}{t} $ :

(i) the local degrees of freedom that would have had higher eigenfrequencies $\vert \Omega \vert > \Omega_t$
have been converted into Local Integrals of Motions (LIOMs) via the projectors $P_{1,2}$ that commute with $H_{eff}$,
 i.e. they have converged towards 
their asymptotic state described by the diagonal ensemble of their local Hamiltonian $H_0$,
while the off-diagonal contributions have been time-averaged-out.

(ii) the remaining degrees of freedom that are characterized by renormalized eigenfrequencies $\vert \Omega \vert < \Omega_t$
have not yet converged towards their asymptotic state, since they have not had enough time to oscillate with their eigenfrequency.

The application of the RSRG-t procedure to various models is described in \cite{vosk13,vosk14,huang17}.
As a final remark, let us mention that another type of Nonequilibrium dynamical renormalization group
has been studied in \cite{heyl15,hauke15}.

\subsection{ Non-equilibrium dynamical scaling of observables }

Following the experimental progress in non-equilibrium dynamics of
ultracold-atomic gases in optical
lattices, there are tremendous theoretical efforts aimed at understanding the
time-evolution of certain observables in closed quantum systems after a sudden
or smooth change of Hamiltonian parameters. In a quench process, both the functional
form of the relaxation and the properties of the 
stationary state are of interest. Here we report numerical results obtained about the non-equilibrium relaxation process in random quantum systems, almost exclusively in one dimension. 

Concerning the functional form of the time-dependence of
the entanglement entropy~\cite{chiara06,igloi12,levine12,bardarson12,zhao16},
 the results depends on whether the random quantum system can be described in terms of free fermions or not. If
the system consists of non-interacting fermions - such as the XX-spin
chain with bond disorder or the critical random transverse-field Ising chain - the
dynamical entanglement entropy grows ultraslowly in time as 
\be
 {\cal S}(t) \sim a \ln \ln t\,,
 \label{eq:S_lnln}
\ee
and saturates in a finite system at a value 
\be
   {\cal S}(\ell) \sim b \ln \ell\,,
 \label{eq:S_longt}  
\ee 
where $\ell$ denotes the size of a block in a bipartite system and can be chosen 
to be proportional to the size of the system $L$~\cite{igloi12,zhao16}. Similar scaling forms have been observed for the non-equilibrium relaxation of the full counting statistics in a disordered free-fermion system \cite{levine12}. 
By the strong disorder RSRG-t  method \cite{vosk13,vosk14} of Sec.\ref{sec:RSRG-t},
 the ratio
of the prefactors in (\ref{eq:S_lnln}) and (\ref{eq:S_longt}) is
predicted as $b/a=\psi_{\mathrm{ne}}$, where $\psi_{\mathrm{ne}}=1/2$ is a critical exponent
in the non-equilibrium process and describes the relation between
time-scale and length-scale as %
\be
\ln t \sim L^{\psi_{\mathrm{ne}}}\;.
\label{psi_ne}
\ee
Numerical estimates of $b/a$ are somewhat larger, being in the range $0.69-0.59$. This discrepancy may be due to the fact that for disordered systems, because of the necessity of calculation of some extremely small eigenvalues, standard
eigenvalue solvers would fail to converge for some large-size samples, 
leading to significant numerical errors.

For interacting fermion models in the Many-Body-Localized phase, the
time-dependence of the dynamical entropy is ${\cal S}(t) \sim  \ln^{\omega} t$
with $\omega \ge 1$, while the saturation value follows the volume law, 
${\cal S}(\ell) \sim \ell$~\cite{bardarson12}. In this case SDRG theory and numerical results are mainly consistent.

The time evolution of the average magnetization, $\overline{m}(t)$, of
the one-dimensional RTIM after global quenches is studied numerically by using multiple precision arithmetic \cite{roosz17}. In this way, the numerical inaccuracies observed in the computation of the entanglement entropy are circumvented.
Starting from a fully ordered initial state,
the relaxation to the critical point is logarithmically slow described by  $\overline{m}(t) \sim \ln^{a'} t$,
and in a finite sample of length $L$, the average magnetization saturates at a size-dependent plateau 
$\overline{m}_p(L) \sim L^{-b'}$; here the two exponents satisfy the relation
$b'/a'=\psi_{\mathrm{ne}}=1/2$. This result is consistent with the SDRG prediction. Starting from a fully disordered initial state,
the magnetization stays at zero for a period of time until $t=t_d$ with $\ln t_d \sim L^{\psi_{\mathrm{ne}}}$ 
and then starts to increase until it saturates to an asymptotic value $\overline{m}_p(L) \sim L^{-b''}$, 
with $b''\approx 1.5$. The distribution of long-time limiting values of the magnetization
shows that the typical and the average values scale differently
and the average is governed by rare events.

For the random Heisenberg chain,
the dynamical properties at finite temperature have been studied numerically and compared with SDRG predictions in
 \cite{herbrych13,shu17}.
The non-equilibrium quench dynamics in quantum spin chains with aperiodic interactions have been studied numerically in \cite{igloi13,roosz14,divakaran18}.

Besides the quantum quenches discussed up to now,
the opposite limit of adiabatic changes of the parameters of the Hamiltonian
has also been studied recently via SDRG to analyze the Kibble-Zurek dynamics through the critical point \cite{mason17}.

\subsection{  Comparison with other RG procedures existing in the field of Many-Body-Localization  }

Since the purpose of the RSRG-X and RSRG-t procedures is to produce an extensive number of Local Integrals of Motion (LIOMS),
it is clear that their validity is limited to Many-Body-Localized Phases :
they allow to analyse the long-ranged order of the excited eigenstates and to study
the phase transitions between different Many-Body-Localized phases.
To analyze the MBL-transition towards the ergodic delocalized phase,
various other RG procedures have been introduced,
such as the Aoki exact RG procedure in configuration space \cite{monthus10mblaoki},
and real-space RG procedures based on entanglement \cite {vosk15mbltransition} 
or resonances \cite{potter15mbltransition,dumitrescu17},
while the phenomenological RG procedure based on the the decomposition into insulating and thermal blocks \cite{zhang16}
is related to some coarsening models that can be exactly solved by Strong Disorder RG (see Appendix E of the review \cite{igloi05review}). An exactly-solvable generalization that takes into account the asymmetry between insulating and thermal blocks corresponds to some Kosterlitz-Thouless scenario \cite{goremykina}.
Finally, the Wegner-RG flow \cite{kehrein06} or variants thereof have been applied recently to various MBL models
\cite{rademaker16,monthus16,pekker17wegner,savitz17,thomson18}.

\section{  Floquet dynamics of periodically driven chains in their localized phases }

\label{sec:Floquet}

The Floquet dynamics of periodically driven quantum systems has attracted a lot of attention recently (see the reviews \cite{bukov15,moessner17}).
The stroboscopic dynamics can be analyzed via the diagonalization of the 
time-evolution-operator over one period $T$
\begin{eqnarray}
 U(T,0) \equiv {\cal T} \left( e^{-i \int_0^T d t H(t) }  \right) =  \sum_{n=1}^{\cal N}  e^{-i \theta_n}   \vert u_n \rangle \langle  u_n \vert
\label{Utimeorder}
\end{eqnarray}
The phases $\theta_n \in ]-\pi,+\pi]$ characterize the eigenvalues $e^{-i \theta_n}  $ of this unitary operator,
while the $ \vert u_n \rangle$ are the corresponding eigenvectors.

To have an explicit evolution operator (instead of the implicit time ordering of Eq \ref{Utimeorder}), it is convenient to consider 
periodic switching between two Hamiltonians $H_0$ and $H_1$ during $T_0$ and $T_1$ respectively
\begin{eqnarray}
 U(T=T_1+T_0,0) = e^{-i T_1 H_1 } e^{-i T_0 H_0 } 
\label{Ucycle}
\end{eqnarray}
The simplest example is
\begin{eqnarray}
H(0 \leq t \leq T_0)  && =H_0 \equiv  - \sum_{n=1}^{N-1} J_n \sigma^z_n \sigma^z_{n+1}
\nonumber \\
H(T_0 \leq t \leq T=T_0+T_1)  && =H_1 \equiv  - \sum_{n=1}^N  h_n \sigma_n^x
\label{h0h1}
\end{eqnarray}
since the time-averaged Hamiltonian is the random transverse field Ising chain.
The RSRG-X rules for the corresponding Floquet evolution operator of Eq. \ref{Ucycle} are analyzed in \cite{monthus17floquet}
and can be considered as a direct generalization of the Fisher RG rules.

More generally, the phase-transitions between different Floquet-Localized-phases are expected to be controlled by Infinite-Disorder-Fixed-Points that can be sudied via SDRG \cite{vasseur18,berdanier}.


\section{ Open dissipative quantum spin chains}

\label{sec:open}

 In the field of open quantum systems,
the interplay between quantum coherence and dissipation
can be analyzed within various frameworks \cite{weiss99,breuer02}.

\subsection{ Quantum spin chains coupled to a bath of quantum oscillators }

The dynamics of a single two-level system coupled to a bath of quantum oscillators 
is the famous 'spin-boson model' \cite{leggett87}.
The generalization for the random quantum Ising chain is described by the Hamiltonian
\begin{eqnarray}
H^{spins} && = - \sum_i h_i \sigma_i^x - \sum_{<i,j>} J_{ij} \sigma_i^z \sigma_j^z
\nonumber \\
H^{bosons} && = \sum_i \sum_k \omega_{i,k} \left( a_{i,k}^{\dagger} a_{i,k} + \frac{1}{2} \right)
\nonumber \\
H^{coupling} && = \sum_i \sigma_i^z \sum_k \lambda_{i,k} \left( a_{i,k}^{\dagger} + a_{i,k}  \right)
\label{hspinsbosons}
\end{eqnarray}
where each spin is coupled to its own local bath of oscillators described by its spectral density
\begin{eqnarray}
{\cal E}_i(\omega) = \pi \sum_k \lambda_{i,k}^2 \delta(\omega- \omega_{i,k}) \oppropto_{\omega \to 0} \omega^{s}
\label{spectraldos}
\end{eqnarray}
The case $s=1$ is called Ohmic dissipation, while $s>1$ is called super-Ohmic and $s<1$ is called sub-Ohmic.

To obtain the appropriate SDRG rules \cite{schehr06,schehr08,hoyos08dissipation,hoyos08smeared,vojta10,hoyos12},
 the idea is to supplement the Fisher SDRG rules of the chain
by the adiabation renormalization of the bath developped for the spin-boson model \cite{leggett87}.
The main conclusion is the smearing of the quantum phase transition by the dissipation for Ohmic and sub-Ohmic baths, with freezing of large magnetic clusters, while super-Ohmic is irrelevant. These predictions have been tested via Monte-Carlo simulations \cite{ali13}.
Related studies based on Landau-Ginzburg functionals can be found in \cite{hoyos07LG,vojta09LG,vojta11LG}.

\subsection{ Lindblad dynamics for random quantum spin chains }

Another popular description of open quantum systems \cite{weiss99,breuer02}
is the Lindblad dynamics for the density matrix $\rho$
\begin{eqnarray}
\frac{\partial \rho }{\partial t} =  -i [H,\rho ]  +  {\cal D}[\rho]
\label{dynlindblad}
\end{eqnarray}
where the unitary dynamics governed by the Hamiltonian $H$ is supplemented 
by the dissipative contribution defined in terms of some set of operators $L_{\alpha}$ 
that describe the interaction with the reservoirs
\begin{eqnarray}
 {\cal D}[\rho] = \sum_{\alpha} \gamma_\alpha \left(L_\alpha \rho L_\alpha^{\dagger} - \frac{1}{2} L_\alpha^{\dagger}L_\alpha \rho- \frac{1}{2} \rho L_\alpha^{\dagger}L_\alpha \right)
\label{dissi}
\end{eqnarray}
In the field of quantum spin chains, it is interesting to consider two reservoirs 
acting only on the two boundary spins in order to impose a current-carrying Non-Equilibrium Steady-State (NESS).

For the XX chain with random fields, a strong disorder boundary renormalization has been introduced 
\cite{monthus17step},
 in order to describe the strong hierarchy of relaxation times as a function of the distance to the boundaries,
and to compute explicitely the sample-dependent step-profile of the magnetization.
This step profile is expected in other localized chains \cite{roeck17}, in stark contrast with the usual linear profile
for diffusive dynamics.
Note however that the addition of dephasing acting on all spins of the bulk destroys the phase coherence 
responsible for the localization properties, 
and produces an effective dynamics described by a classical exclusion process with randomness \cite{monthus17dephasing}.
Another study concerning the Lindblad dynamics 
with a contact between random and pure quantum XX spin chains can be found in \cite{chatelain17}.

\section{ Anderson localization models }

\label{sec_AndersonLoc}

In the field of Anderson localization (see the review \cite{evers08}), one is interested into the localization/delocalization
properties of the eigenstates of tight-binding Hamiltonian of the form
\begin{eqnarray}
H = \sum_i H_{ii} \vert i> < i \vert + \sum_{i \ne j} H_{ij} (\vert i> < j \vert + \vert j> < i \vert )
\label{handerson}
\end{eqnarray}
that can be defined for various geometries.

The goal of the SDRG \cite{mard14,mard17} is to analyze the properties at zero-energy $E=0$ corresponding to the middle of the spectrum. 
The iterative elimination of the strongest on-site energy $ H_{ii}$ or the strongest off-diagonal hopping $H_{ij}$
( in absolute value ) leads to the following decimation rules.
The decimation of the on-site energy $\Omega= \vert H_{i_0i_0} \vert$ yields the RG rule (even for $k=p$)
\begin{eqnarray}
H^R_{kp}  && = H_{kj} - \frac{H_{k i_0} H_{i_0 p}}{ H_{i_0i_0} }
\label{andersonrg1}
\end{eqnarray}
while the decimation of the off-diagonal coupling $\Omega= \vert H_{i_0j_0} \vert$ produces the RG rule (even for $k=p$)
\begin{eqnarray}
H^R_{kp}  && = H_{kp} + \frac{ H_{i_0j_0} (H_{k i_0} H_{j_0 p} + H_{k j_0} H_{i_0 p} )
}{ H^2_{i_0j_0} - H_{i_0j_0}  H_{j_0j_0} }
\nonumber \\
&& - \frac{ H_{i_0i_0} H_{k j_0} H_{j_0 p}  +H_{j_0j_0} H_{k i_0} H_{i_0 p}
}{ H^2_{i_0j_0} - H_{i_0j_0}  H_{j_0j_0} }
\label{andersonrg2}
\end{eqnarray}
These SDRG rules actually coincide with the exact Aoki RG rules at zero-energy $E=0$ \cite{aoki80,aoki82,monthus09aoki}.
As a consequence, if one focuses on zero-energy $E=0$, the SDRG rules are exact in both phases (localized and delocalized)
as well as at the Anderson phase transition between them.

The SDRG rules have been applied to analyze the localization properties in $d=1$  \cite{mard14} and $d=2$ \cite{monthus09aoki}
and to characterize the critical properties of the Anderson delocalization transition in $d=3$ \cite{mard17,monthus09aoki,tarquini17},
while the application in higher dimensions $d>3$ \cite{mard17,tarquini17}
points towards an infinite upper critical dimension $d_{upper}=+\infty$ for the Anderson transition.
Note that another SDRG rules based on the Inverse Participation Ratios of eigenstates have been proposed in \cite{johri14},
while the effects of rare resonances on various observables is discussed in \cite{bhatt12}.

As a final remark, let us mention that several other real-space renormalization approaches have been 
introduced to analyze the multifractality of eigenstates at the localization-delocalization transition
in various models,
in particular the Levitov RG reviewed in \cite{evers08},
the block-RG  \cite{monthus10cascade,monthus11dysonanderson}
and the Wegner flow approach \cite{quito16}.
 
\section{ Random contact process  }

\label{sec:non-equilibrium classical}

The contact process \cite{harris74,liggett05} is a basic model in the fields of 
epidemic spreading and population dynamics. It is
defined on a lattice, the sites of which are either active (infected) or
inactive (healthy). The time evolution is a continuous-time Markov
process with the following independent transition rates. Site $i$, if it
is active, becomes spontaneously inactive with a rate $\mu_i$ or it
activates site $j$, provided the latter is inactive, with a rate
$\lambda_{ij}$. 
This model in the simplest case with homogeneous parameters and nearest neighbour spreading falls into the universality class of
directed percolation.

In experimental realizations of directed percolation, quenched disorder is observed to play an inevitable role, therefore different variants of randomness in the contact process have been studied theoretically. According to the Harris criterion, quenched disorder (both spatial and temporal) is a relevant perturbation, therefore new type of critical bahaviour is expected to be present in these systems.

\subsection{Strong disorder RG rules}

The contact process with random short range interactions (both the $\lambda_{ij}$ and the $\mu_i$ are i.i.d. random variables) has been studied by the SDRG method, the basic results can be found in Review \cite{igloi05review}. The elementary decimation steps are the following. Having a very strong activation rate $\lambda_{ij}=\Omega$, the two connected sites form a cluster in the presence of an effective recovery rate
$\tilde{\mu}=2\mu_i\mu_j/\lambda_{ij}$. On the contrary, for a strong recovery rate $\mu_i=\Omega$, this site is almost always inactive, and there are effective branching rates between all sites $j,k$, which are nearest neighbours to $i$, as $\tilde{\lambda}_{jk}=\lambda_{ji}\lambda_{ik}/\mu_i$. Supplementing this relation with the maximum rule we arrive to the elementary decimation rules which are very similar to that of the RTIM. Indeed, for nearest neighbour interactions and for strong enough initial disorder, the extra prefactor is unimportant, and the critical behaviour is controlled by the IDFP of the RTIM. The infinite-randomness scenario has been checked by large scale Monte-Carlo simulations in $d=2$ \cite{vojta08mc,vojta09} and in $d=3$ \cite{vojta12}.
For weaker disorder, the cross-over between the weak- and the strong disorder scaling regions
is analyzed in \cite{hoyos08}. 
Detailed results on the distribution of dynamical observables can be found in \cite{juhasz13dyn}.
The contact process with asymmetric spreading has also been studied by SDRG \cite{juhasz13a}.
In dimension $d=5$, the Griffiths singularities are analyzed in \cite{vojta14five},
as an example where they can co-exist with a clean critical behavior predicted by the Harris criterion \cite{vojta14rareharris}.
The effects of long-ranged correlated disorder is studied in \cite{ibrahim14}.
The contact process on aperiodic chains (instead of random chains) has been found to display double-logarithmic periodic oscillations via real-space renormalization \cite{barghathi14}.
The numerical study of the contact process on complex networks has revealed the importance of Griffiths phases
and other rare region effects as a consequence of topological heterogeneity of the network \cite{munoz10complex}.
Finally, the effect on random-field disorder on the Generalized contact process has been studied in \cite{vojta12and15}.

\subsection{Long range spreading}

Spreading of epidemics with long range infections - which has a power-law distribution - can happen in different situations. This type of process can be modelled by the contact process in which the activation rates are parametrised as:
\be 
\lambda_{ij}=\Lambda_{ij}r_{ij}^{-(d+\sigma)},
\label{lambda}
\ee
where $r_{ij}$ is the Euclidean distance between site $i$ and $j$,
and $\Lambda_{ij}$ are $O(1)$ i.i.d. quenched random variables, while the recovery rates $\mu_i$ are
also i.i.d. quenched random variables as before.

The SDRG trajectories have been analysed \cite{juhasz15} in the same way as that of the RTIM with LR interaction of Sec.\ref{sec:LR}. Analytical solution of the primary model in 1D, as well as numerical implementation of the renormalization with the maximum rule in 1D and 2D lead to identical critical scaling behavior as illustrated in Fig.\ref{fig_2}. In the language of the contact process, the following consequences have been obtained.
Starting from a single infected site, the average survival probability is found
to decay as $P(t) \sim t^{-d/z}$ up to multiplicative logarithmic corrections.
Below the epidemic threshold, a Griffiths phase emerges, where the dynamical
exponent $z$ varies continuously with the control parameter and tends to
$z_c=d+\sigma$ as the threshold is approached. 
At the threshold,  the spatial extension of the infected cluster (in surviving
trials) is found to grow as $R(t) \sim t^{1/z_c}$ with a multiplicative
logarithmic correction, and the average number of infected sites in surviving
trials is found to increase as $N_s(t) \sim (\ln t)^{\chi}$ with $\chi=2$ in one
dimension. These results have been confirmed by numerical Monte Carlo simulations \cite{juhasz15}.
We note that on a long-range connected network, the contact process has infinite disorder criticality \cite{juhasz13b}.

\subsection{Temporal disorder}

The contact process in time-varying environmental noise, i.e. temporal disorder,
has been considered in Refs. \cite{vazquez11,vojta15,barghathi16}. The system is spatially homogeneous, but the (nearest neighbour) activation and recovery rates are time dependent:
\be
\lambda(t)=\lambda_n,\quad \mu(t)=\mu_n \quad (t_n<t<t_{n+1})\;.
\ee
\textbf{In the mean-field approximation} the time evolution of the density $\rho$ of the active sites follows the differential equation:
\be
\dot{\rho}(t)=\left[\lambda(t)-\mu(t)\right]\rho(t)-\lambda(t) \rho^2(t)\;,
\ee
the solution of which in the interval $t_n<t<t_{n+1}$ for a given disorder realization is given by:
\be
\rho_{n+1}^{-1}=a_n \rho_{n}^{-1} +c_n\;.
\ee
Here $\rho_n=\rho(t_n)$, $a_n=\exp[(\mu_n-\lambda_n)\Delta t]$ and the constant $c_n=(a_n-1)\lambda_n/(\mu_n-\lambda_n)$.

The strong disorder (or strong noise) RG consists of iteratively decimating the weakest spreading and recovery segments, characterised by $a_i^{\uparrow}>1$ and $a_i^{\downarrow}<1$, respectively and $\Omega=\min(a_i^{\uparrow},1/a_i^{\downarrow})$. The decimation equations are given by: $\tilde{a}^{\uparrow}=a_{i+1}^{\uparrow}a_i^{\uparrow}/\Omega$
(for $\Omega=1/a_i^{\downarrow}$) and $1/\tilde{a}^{\downarrow}=(1/a_i^{\downarrow})(1/a_{i-1}^{\downarrow})/\Omega$ (for $\Omega=a_i^{\uparrow}$), which are equivalent to those of the RTIM in one dimension. Thus the critical behavior in the mean-field approximation is controlled by the IDFP of this model.

\textbf{In finite dimensions} the decimation equations are $\tilde{a}^{\uparrow}=a_{i+1}^{\uparrow}a_i^{\uparrow}/\Omega$
(for $\Omega=1/a_i^{\downarrow}$) and $(1/\tilde{a}^{\downarrow})^{1/D}=(1/a_i^{\downarrow})^{1/D}+(1/a_{i-1}^{\downarrow})^{1/D} -\Omega^{1/D}$ (for $\Omega=a_i^{\uparrow}$). Here in the second equation, one takes into account that in finite dimensions by decimating $\Omega=a_i^{\uparrow}$ the radii of the combined active clusters grows linearly in time. The elementary RG steps in this case are equivalent to that of the RTIM in one dimension with long-range interactions, see in Sec.\ref{sec:primary} and the RG trajectories are illustrated in Fig.\ref{fig_2}. The singular properties of the observables in the contact process can be found in \cite{vojta15,barghathi16} and numerical calculations are performed in Ref. \cite{barghathi16}. 

Temporal disorder at first-order non-equilibrium phase transitions has been studied in \cite{fiore18}.
More generally, the effect of spatio-temporal disorder on various equilibrium and nonequilibrium critical points
is discussed in \cite{vojta16spatiotemp}.


\section{ Classical master equations  }

\label{sec:mastereq}

The stochastic dynamics in random classical models usually displays 
a broad continuum of relevant time scales.
The scaling between the characteristic time $t$ and the linear length $L$ 
can be either activated with some exponent $\psi$ characterizing some Infinite Disorder Fixed Point
\begin{eqnarray}
\ln t = L^{\psi}  
\label{activated}
\end{eqnarray}
or power-law with some dynamical exponent $z$ that may vary continuously as a function of the model parameters
\begin{eqnarray}
t = L^{z} 
\label{powerz}
\end{eqnarray}
The limit $z \to \infty$ corresponds to the activated scaling (Eq \ref{activated}) of the Infinite Disorder Fixed Point,
but the whole region 
\begin{eqnarray}
1<z<+\infty
\label{regionz}
\end{eqnarray}
can usually be well described by the Strong Disorder approximation.

\subsection{  Real-space renormalization for random walks in random media }

As reviewed in \cite{igloi05review}, Strong Disorder RG procedures have been applied to various models of random walks 
in random media. The main principle is the iterative elimination of the fastest degree of freedom to obtain the effective dynamics for the slowest ones.
In this field, new developments since 2005 include 
one-dimensional random walks with dilute absorbers \cite{ledoussal09}
or with long-range connections \cite{juhasz12RWLR},
random walks on strips \cite{juhasz08RWstrip,juhasz10channel} and on arbitrary networks \cite{juhasz12RWnetwork},
and in two-dimensional self-affine random potentials \cite{monthus10RWaffine}.

\subsection{  RG in configuration-space for the dynamics of classical many-body models  }

The real-space SDRG for random walks in random media has been generalized into the configuration-space SDRG \cite{monthus08broad}
for any classical master equation governing the dynamics of the probability $P_t({\cal C})$ to be in configuration ${\cal C} $
at time $t$
\begin{eqnarray}
\frac{ dP_t \left({\cal C} \right) }{dt}
= \sum_{\cal C '} P_t \left({\cal C}' \right) 
W \left({\cal C}' \to  {\cal C}  \right) 
 -  P_t \left({\cal C} \right) W_{out} \left( {\cal C} \right)
\label{master}
\end{eqnarray}
where $ W \left({\cal C}' \to  {\cal C}  \right) $ 
represents the transition rate per unit time from configuration 
${\cal C}'$ to ${\cal C}$ while
\begin{eqnarray}
W_{out} \left( {\cal C} \right)  \equiv
 \sum_{ {\cal C} '} W \left({\cal C} \to  {\cal C}' \right) 
\label{wcout}
\end{eqnarray}
denotes the total exit rate out of configuration ${\cal C}$.
The SDRG rule consists in the elimination of the configuration $ {\cal C}^*$ with the highest total exit rate $ W_{out}(  {\cal C}^*)$ to obtain the new renormalized transition rates between surviving configuration
\begin{eqnarray} 
 W^{new}({\cal C}_i && \to {\cal C}_j )  =W^{old}({\cal C}_i \to {\cal C}_j ) 
\nonumber \\
&& + W^{old}({\cal C}_i \to {\cal C}^* )
\frac{W^{old}({\cal C}^* \to {\cal C}_j )}
{W_{out}(  {\cal C}^*) }
\label{wijnew}
\end{eqnarray}
and the new exit rates
\begin{eqnarray}
  W^{new}_{out}({\cal C}_i) && = W^{old}_{out}({\cal C}_i)  
\nonumber \\
&&  - W^{old}({\cal C}_i \to {\cal C}^* ) 
\frac{ W^{old}({\cal C}^* \to {\cal C}_i )}{W_{out}(  {\cal C}^*)}
 \label{wioutnew}
\end{eqnarray}
 The physical interpretation of this procedure is as follows :
the time spent in the decimated configuration ${\cal C}^*$ is neglected
with respects to the other time scales remaining in the system; 
the remaining configurations represents some 'valleys' in configuration space
that takes into account all the previously decimated configurations.
As a consequence of the multiplicative structure
of the renormalization rule of Eq \ref{wijnew},
the renormalized rates $W({\cal C} \to {\cal C}' )$
can rapidly become very small and the appropriate variables 
are the logarithms of the transition rates, called 'barriers'
\begin{eqnarray}
B ({\cal C} \to {\cal C}' ) \equiv - \ln W({\cal C} \to {\cal C}' ) 
\label{defbout}
\end{eqnarray}

This SDRG procedure has been applied numerically to 
interfaces in two-dimensional random media \cite{monthus08broad,monthus08flow,monthus08valley}
with possibly driving \cite{monthus08interface}.
Note that the idea to eliminate fast degrees of freedom in classical master equations
is very natural and has been thus developed independently 
 in many other contexts (see the recent review \cite{Bo17} and references therein).
The SDRG rules above are actually similar to the exact RG rules concerning first-passage times \cite{monthus10first}
where the application to spin-glasses is discussed.

As a final remark, it is important to stress that for many-body classical models,
the fact that the above RG rules are defined in configuration space clearly limits the numerical implementation to 
small sizes. As a consequence, various other types of real-space RG procedures have been developed for 
the dynamics of classical spin models, 
in particular boundary-RG for the dynamics in $d=1$ \cite{monthus13zerotemperature}
and on the Cayley tree \cite{monthus13cayley}
or block-RG for for the dynamics of Long-Ranged ferromagnetic \cite{monthus13dyson} or Spin-Glass models \cite{monthus14LRSG,monthus16dysonSG}.


\section{ Random classical oscillators }

\label{sec:oscillators}

\subsection { Random elastic networks }

The model of random masses $m_i$ connected by random springs $K_{ij}$ is one of the oldest problem in the field of localization of classical disordered models \cite{dyson53}. One is interested into the Newton equations of motions for the displacements $u_i(t)$
\begin{eqnarray}
m_i \frac{d^2 u_i}{dt^2} =\sum_j K_{ij} ( u_j-u_i)
\label{newton}
\end{eqnarray}
for various geometries.
The issue of the localization properties of the eigenmodes (phonon localization) 
is related to the Anderson localization properties of tight-binding models discussed in Section \ref{sec_AndersonLoc}
even if they are some differences (see for instance \cite{monthus10phonon} and references therein).

Various slightly different SDRG rules have been proposed :
the idea is to eliminate iteratively either
only the masses \cite{hastings03}, or only the couplings \cite{amir10},
or both \cite{monthus10elastic} as we now describe.
One first needs to identify the local degree of freedom oscillating with the highest frequency.

The frequency $\Omega_{i,j} $ associated to the spring $K_{i,j}$ between two masses $m_i$ and $m_j$ is defined by \cite{monthus10elastic}
\begin{eqnarray}
\Omega_{i,j}^2 \equiv  K_{i,j} \left(\frac{1}{m_i}+ \frac{1}{m_j} \right) 
\label{omegaij}
\end{eqnarray}
while the frequency $\Omega_i$ associated to the mass $m_i$ connected to the springs $K_{ij}$ is given by \cite{monthus10elastic}
\begin{eqnarray}
\Omega_i^2 \equiv \frac{1 }{m_i} \sum_{j} K_{i,j}
\label{omegai}
\end{eqnarray}

The renormalization scale $\Omega$ is defined as the 
 highest local frequency remaining in the system
among all the frequencies associated with masses or spring constants.
\begin{eqnarray}
\Omega \equiv {\rm max}  \{\Omega_i , \Omega_{i,j} \} 
\label{rgscale}
\end{eqnarray}

 If the highest frequency $\Omega=\Omega_{i_0,j_0}$
is associated with the spring constant $K_{i_0,j_0}$,
the two masses $m_{i_0}$ and $m_{j_0}$ are replaced by their center of mass $ G(i_0,j_0)$
of mass 
\begin{eqnarray}
m_{G(i_0,j_0)}= m_{i_0} + m_{j_0}
\label{Gi0j0}
\end{eqnarray}
and the spring constants are replaced by
spring constants linked to their center of mass
\begin{eqnarray}
K_{j,G(i_0,j_0)}= K_{j,i_0}+K_{j,j_0}
\label{KjGi0j0}
\end{eqnarray}
This renormalization step thus constructs a cluster of strongly-coupled masses oscillating together.

If the highest frequency $\Omega=\Omega_{i_0}$
is associated with the mass $i_0$, 
the mass $m_{i_0}$ is eliminated, and the spring constants are renormalized according to
\begin{eqnarray}
K_{i,j}^{new} = K_{i,j}+ \frac{K_{i,i_0} K_{i_0,j} }
{\sum_{n} K_{i_0,n}} 
\label{Kijnew}
\end{eqnarray}
This renormalization step thus constructs an isolated localized oscillating mode.

These SDRG rules coincide with the exact Aoki RG rules at zero frequency $\omega=0$ \cite{monthus10elastic},
so they are expected to remain a good approximation at low frequency.

The SDRG approach has been applied to complex networks to analyze their localization properties \cite{hastings03}
as well as to some matrix model in relation with slow relaxation in glasses  \cite{amir10}.

\subsection{ Synchronisation of interacting non-linear dissipative classical oscillators }

In the field of emergent collective structures in nonequilibrium systems, 
the spontaneous synchronization of interacting nonlinear oscillators 
is one of the most studied phenomenon \cite{strogatz00,pikovsky01,pikovsky15}.
Each oscillator is characterized by its mass $m_i$ and its own frequency $\omega_i$,
while the interactions between oscillators are described by couplings $K_{ij}$ that define the geometry of the network of oscillators.
The dynamical equations for the phases $\theta_i(t)$ of the oscillators are written in the dissipative limit (first order in time)
\begin{eqnarray}
m_i \frac{d \theta_i}{dt} =m_i \omega_i +\sum_j K_{ij} \sin( \theta_j-\theta_i)
\label{kuramoto}
\end{eqnarray}

The aim of the SDRG procedure \cite{kogan09,lee09} is to construct clusters of frequency-synchronized-oscillators.
The two decimation possibilities are as follows.
The decimation of a coupling $K_{ij}$ corresponds to the synchronization of the two corresponding oscillators
and its replacement by a single renormalized oscillator.
The decimation of a frequency $\omega_i$ means that the corresponding oscillator rotates freely and does not contribute to the global synchronization. 
The details of the SDRG rules and the numerical results are described in \cite{kogan09,lee09}.

\section{ Other classical models }

\label{sec:otherclassical}

\subsection{ Equilibrium properties of random systems} 

SDRG has been also used to analyze the equilibrium phase transitions of various classical systems,
as reviewed in \cite{igloi05review}, while more recent applications include
the randomly layered Heisenberg magnet \cite{mohan10},
 the wetting transition on the Cayley tree \cite{monthus09wetting},
the DNA denaturation transition \cite{monthus17dna}.
In the field of classical spin-glasses, some SDRG procedure have been also introduced
to study the spin-glass phase of the Long-Ranged Spin-Glass chain \cite{monthus14LRSGzero}
or the fractal dimension of interfaces in Short-Ranged Spin-Glasses as a function of the dimension $d$ 
 \cite{monthus15fractal,wang17,wang18}.

\subsection{ Extremes of stochastic processes}

As explained in detail in the Review \cite{igloi05review}, SDRG procedures are closely related to 
the statistics of extrema of some random processes associated to the disorder variables : 
for instance, the Fisher solution is directly related to the statistics of extrema of the Brownian motion\cite{juhasz06}.
Reciprocally, the Extreme Value Statistics of various stochastic processes
can be analyzed via SDRG \cite{schehr10}
(while the Extreme Value Statistics of independent variables is analyzed via RG in the series of works \cite{extreme}).
Some coagulation model with extremal dynamics has been also studied \cite{juhasz09extremal},
in relation with previous works reviewed in Appendix E of \cite{igloi05review}).


\section{ Conclusion  }

In summary, we have reviewed the new developments of Strong Disorder RG methods since 2005.
For the quantum phase transitions of ground states, the critical properties
have been described for short-ranged models in higher dimensions $d>1$ and for long-ranged models.
The scaling of the entanglement entropy has been discussed both for critical ground-states and after quantum quenches.
In Many-Body-Localized phases, we have explained how the SDRG procedure has been extended into RSRG-X procedure to construct the whole set excited stated and into the RSRG-t procedure for the unitary dynamics. Other generalizations of the SDRG approach concern non-isolated quantum models, namely periodically driven models (Floquet dynamics) or dissipative models (coupling to external baths).
 We have then focused on the recent progress for classical disordered models, with the contact process for epidemic spreading, the strong disorder renormalization procedure for general master equations, the localization properties of random elastic networks and the synchronization of interacting non-linear dissipative oscillators.

In conclusion, SDRG methods have flourished over the years well beyond their initial scopes, and 
we thus expect that they will continue to be developed even further in the future.

\begin{acknowledgments}
It is a pleasure to thank collaborations and discussions with several colleagues: F. Alet, E. Altman, G. Biroli, P. Calabrese, J. Cardy, C. Chatelain, L. Cugliandolo, U. Divakaran, D. Fisher, T. Garel, J.A. Hoyos, D. Huse, R. Juh\'asz, I.A. Kov\'acs, N. Laflorencie, P. Lajk\'o, P. Le Doussal, Y.-C. Lin, J. M. Luck, R. M\'elin, J. E. Moore, M.A. Moore, G. \'Odor, F. Pollmann, G. Refael, H. Rieger, G. Ro\'osz, A. Sandvik, L. Santen, G. Schehr, M. Schiro, Zs. Szatm\'ari, L. Turban, R. Vasseur, T. Vojta and Z. Zimbor\'as.

The work of F.I. has been supported by the Hungarian Scientific Research Fund under Grants
No. K109577, No. K115959 and No. KKP-126749. He thanks the IPhT Saclay for hospitality.

\end{acknowledgments}
\begin{center}
\textbf{Author contribution statement:}
\end{center}
F.I. and C.M. have contributed equally by planning the structure of the review and by writing the text.



\begin{thebibliography}{99}



\bibitem{ma79}       
       S.-K. Ma, C. Dasgupta, and C.-k. Hu,
Random Antiferromagnetic Chain,
       Phys. Rev. Lett. 43, 1434 (1979) ; \\
      C. Dasgupta and S.-K. Ma
Low-temperature properties of the random Heisenberg antiferromagnetic chain
       Phys. Rev. B 22, 1305 (1980).

\bibitem{fisher92}
D. S. Fisher,
Random transverse field Ising spin chains,
Phys. Rev. Lett. 69, 534 (1992) ; \\
D. S. Fisher,
Critical behavior of random transverse-field Ising spin chains
Phys. Rev. B 51, 6411 (1995).



\bibitem{igloi05review}
F. Igl\'oi and C. Monthus, 
Strong disorder RG approach of random systems,
Phys. Rep. 412, 277 (2005).


 
 \bibitem{kovacs09} 
 I. A. Kov\'acs and F. Igl\'oi, 
Critical behavior and entanglement of the random transverse-field Ising model between one and two dimensions,
Phys. Rev. B 80, 214416 (2009).
 
  
 \bibitem{kovacs10} 
 I. A. Kov\'acs and F. Igl\'oi, 
Renormalization group study of the two-dimensional random transverse-field Ising model,
Phys. Rev. B 82, 054437 (2010).
 
 \bibitem{kovacs11a} 
 I. A. Kov\'acs and F. Igl\'oi, 
Infinite disorder scaling of random quantum magnets in three and higher dimensions,
 Phys. Rev. B 83, 174207 (2011).

 \bibitem{kovacs11b} 
 I. A. Kov\'acs and F. Igl\'oi, 
Renormalization group study of random quantum magnets,
 J. Phys.: Condens. Matter 23, 404204 (2011).

 \bibitem{iyer12} 
 S. Iyer, D. Pekker and G. Refael, 
A Mott Glass to Superfluid Transition for Random Bosons in Two Dimensions
Phys. Rev. B 85, 094202 (2012).



\bibitem{vojta06}
T. Vojta  
Rare region effects at classical, quantum and nonequilibrium phase transitions
 J. Phys. A: Math. Gen. 39 R143 (2006).

\bibitem{vojta13}
T. Vojta,
Phases and phase transitions in disordered quantum systems,
arxiv: 1301.7746.




\bibitem{garnerone09}
S. Garnerone, N. Jacobson, S. Haas and P. Zanardi,
Fidelity Approach to the Disordered Quantum XY Model
Phys. Rev. Lett. 102 057205 (2009);
N. T. Jacobson, S. Garnerone, S. Haas and P. Zanardi,
Scaling of the fidelity susceptibility in a disordered quantum spin chain
Phys. Rev. B 79, 184427 (2009).



\bibitem{monthus15entropygs}
C. Monthus, 
Pure and Random Quantum Ising Chain : Shannon and R\'enyi entropies of the ground state via real space renormalization
J. Stat. Mech.  P04007 (2015).


 
 
 \bibitem{kovacs13} 
 I. A. Kov\'acs and F. Igl\'oi, 
Boundary critical phenomena of the random transverse Ising model in $D\ge 2$ dimensions,
 Phys. Rev. B 87, 024204 (2013).

 \bibitem{juhasz14star} 
R. Juh\'asz, 
Critical behavior of models with infinite disorder at a star junction of chains,
	J. Stat. Mech. P08005 (2014)

 \bibitem{monthus15star} 
C. Monthus,
Star junctions and watermelons of pure or random quantum Ising chains : finite-size properties of the energy gap at criticality, 
J. Stat. Mech. P06036 (2015)

 \bibitem{igloi18} 
F. Igl\'oi and I. A. Kov\'acs,
Transverse-spin correlations of the random transverse-field Ising model,  
Phys. Rev. B 97, 094205 (2018)

\bibitem{galambos78}
 J. Galambos, 
The Asymptotic Theory of Extreme Order Statistics (Wiley, New York, 1978).

\bibitem{juhasz06} 
R.  Juh\'asz, Y-C. Lin, F. Igl\'oi, 
Strong Griffiths singularities in random systems and their relation to extreme value statistics, 
Phys. Rev. B 73, 224206 (2006)

\bibitem{lin07} 
Y-C. Lin, F. Igl\'oi and H. Rieger,
Entanglement entropy at infinite randomness fixed points in higher dimensions, 
 Phys. Rev. Lett. 99, 147202 (2007)
  
\bibitem{yu08} 
R. Yu, H. Saleur and S. Haas,
Entanglement Entropy in the Two-Dimensional Random Transverse Field Ising Model,
 Phys. Rev. B 77, 140402(R) (2008)

\bibitem{laumann12} 
C.R. Laumann, D.A. Huse, A.W.W. Ludwig, G. Refael, S. Trebst and M. Troyer, 
Strong-disorder renormalization for interacting non-Abelian anyon systems in two dimensions,
Phys. Rev. B 85, 224201 (2012)

\bibitem{dimitrova11}
O. Dimitrova and M. M\'ezard,
The cavity method for quantum disordered systems: from transverse random field ferromagnets to directed polymers in random media,
 J. Stat. Mech. P01020 (2011).

\bibitem{monthus_garel1} 
C. Monthus and T. Garel, 
Random Transverse Field Ising Model in dimension $d>1$ : scaling analysis in the disordered phase from the Directed Polymer model,
J. Phys. A: Math. Theor. 45, 095002 (2012).

\bibitem{monthus_surface} 
C. Monthus and T. Garel,
Random Transverse Field Ising Model in dimension $d=2,3$ : Infinite Disorder scaling via a non-linear transfer approach
 J. Stat. Mech. P01008 (2012).

\bibitem{monthus_garel2} 
C. Monthus and T. Garel, 
Strong Disorder RG principles within a fixed cell-size real space renormalization : application to the Random Transverse Field Ising model on various fractal lattices,
J. Stat. Mech. P05002 (2012).

\bibitem{monthus_garel3} 
C. Monthus and T. Garel, 
Random Transverse Field Ising model on the Cayley Tree : analysis via Boundary Strong Disorder Renormalization,
J. Stat. Mech. P10010 (2012).

\bibitem{monthus_garel4} 
C. Monthus and T. Garel, 
Random Transverse Field Ising model in $d=2$ : analysis via Boundary Strong Disorder Renormalization,
J. Stat. Mech. P09016 (2012).

\bibitem{miyazaki12} 
R. Miyazaki and H. Nishimori,
Real-space renormalization-group approach to the random transverse-field Ising model in finite dimensions,
 Phys. Rev. E 87, 032154 (2013).

\bibitem{ioffe10}
L. B. Ioffe and M. M\'ezard, 
Disorder-Driven Quantum Phase Transitions in Superconductors and Magnets,
Phys. Rev. Lett. 105, 037001 (2010);
M.V. Feigel'man, L. B. Ioffe and M. M\'ezard, 
Superconductor-Insulator transition and energy localization,
Phys. Rev. B 82, 184534 (2010).



\bibitem{dutta15} 
For a review see, A. Dutta, G. Aeppli, B. K. Chakrabarti, U. Divakaran, T. F. Rosenbaum and D. Sen, 
Quantum Phase Transitions in Transverse Field Spin Models: From Statistical Physics to Quantum Information,
 Cambridge University Press, Cambridge, (2015)

\bibitem{friedenauer08} 
A. Friedenauer, H. Schmitz, J. T. Glueckert, D. Porras, and T. Schaetz,
Simulating a quantum magnet with trapped ions, 
Nat. Phys. 4, 757 (2008).

\bibitem{kim10} 
K. Kim, M.-S. Chang, S. Korenblit, R. Islam, E. E. Edwards, J. K. Freericks, G.-D. Lin, L.-M. Duan, and
C. Monroe, 
Quantum simulation of frustrated Ising spins with trapped ions,
Nature (London) 465, 590 (2010).

\bibitem{islam11} 
R. Islam, E. E. Edwards, K. Kim, S. Korenblit, C. Noh,
H. Carmichael, G.-D. Lin, L.-M. Duan, C.-C. J. Wang,
J. Freericks, and C. Monroe, 
Onset of a Quantum Phase Transition with a Trapped Ion Quantum Simulator,
Nat. Commun. \textbf{2}, 377 (2011).

\bibitem{britton12} 
J.W. Britton, B. C. Sawyer, A. C. Keith, C. J. Wang, J. K.
Freericks, H. Uys, M. J. Biercuk, and J. J. Bollinger,
Engineered two-dimensional Ising interactions in a trapped-ion quantum simulator with hundreds of spins,
Nature (London) \textbf{484}, 489 (2012).

\bibitem{islam13} 
R. Islam, C. Senko, W. C. Campbell, S. Korenblit, J. Smith, A.
Lee, E. E. Edwards, C.-C. J. Wang,
J. K. Freericks, and C. Monroe, 
Emergence and Frustration of Magnetism with Variable-Range Interactions in a Quantum Simulator,
Science \textbf{340}, 583 (2013).

\bibitem{juhasz14a}
R. Juh\'asz, I. A. Kov\'acs, and F. Igl\'oi,  
Random transverse-field Ising chain with long-range interactions
Europhys. Lett.  107, 47008 (2014).

\bibitem{altman04}
E. Altman, Y. Kafri, A. Polkovnikov, and G. Refael, 
Phase transition of one dimensional bosons with strong disorder
Phys. Rev. Lett. 93, 150402 (2004)

\bibitem{altman10}
E. Altman, Y. Kafri, A. Polkovnikov, and G. Refael, 
Superfluid-insulator transition of disordered bosons in one-dimension
Phys. Rev. B 81, 174528 (2010).



\bibitem{kovacs16} 
  I. A. Kov\'acs, R. Juh\'asz and F. Igl\'oi, 
Long-range random transverse-field Ising model in three dimensions,
Phys. Rev. B 93, 184203 (2016).




\bibitem{dutta06LRdilute}
A. Dutta and R. Loganayagam,
Effect of long-range connections on an infinite randomness fixed point associated with the quantum phase transitions in a transverse Ising model
Phys. Rev. B 75 052405 (2007);
U. Divakaran and A. Dutta,
Long-range connections, quantum magnets and dilute contact processes
Physica A 384 39 (2007)


\bibitem{monthus15b} 
C. Monthus,
Dyson hierarchical quantum ferromagnetic Ising chain with pure or random transverse fields
 J. Stat. Mech. P05026 (2015).

\bibitem{monthus15c} 
C. Monthus,
Dyson Hierarchical Long-Ranged Quantum Spin-Glass via real-space renormalization
J. Stat. Mech. P10024 (2015).

\bibitem{juhasz14b}
R. Juh\'asz, 
Infinite-disorder critical points of models with stretched exponential interactions,
J. Stat. Mech. P09027 (2014).

 \bibitem{moure15} 
N. Moure, S. Haas, S. Kettemann, 
Many-Body Localization Transition in Random Quantum Spin Chains with Long-Range Interactions,
Europhysics Letters 111, 27003 (2015).

 \bibitem{moure18} 
 N. Moure, H.Y. Lee, S. Haas, R. N. Bhatt and S. Kettemann, 
Disordered Quantum Spin Chains with Long-Range Antiferromagnetic Interactions,
Phys. Rev. B 97, 014206 (2018)
 


\bibitem{vojta10gruneisen}
T. Vojta and J. A. Hoyos, Magnetic Gr\"uneisen ratio of the random transverse-field Ising chain,
phys. stat. sol (b) 247, 525 (2010)



\bibitem{shivamoggi10}
V. Shivamoggi, G. Refael and J. E. Moore, 
Majorana fermion chain at the Quantum Spin Hall edge,
Phys. Rev. B 82, 041405(R) (2010).


\bibitem{lin17} 
Y.P. Lin, Y.J. Kao, P. Chen, Y.C. Lin, 
Griffiths Singularities in the Random Quantum Ising Antiferromagnet: A Tree Tensor Network Renormalization Group Study, 
Phys. Rev. B 96, 064427 (2017).

\bibitem{lajko18} 
P. Lajk\'o and F. Igl\'oi, 
Numerical study of the random quantum Ising antiferromagnetic chain
 (to be published)





\bibitem{goswami08}
P. Goswami, D. Schwab and S. Chakravarty,
Rounding by Disorder of First-Order Quantum Phase Transitions: Emergence of Quantum Critical Points,
Phys. Rev. Lett. 100, 015703 (2008).



\bibitem{hrahsheh12} 
 F. Hrahsheh, J. A. Hoyos and  T. Vojta, 
Rounding of a first-order quantum phase transition to a strong-coupling critical point,
Phys. Rev. B 86, 214204 (2012).

\bibitem{hrahsheh14} 
F. Hrahsheh, R. Narayanan, J. A. Hoyos and T. Vojta, 
Strong-randomness infinite-coupling phase in a random quantum spin chain,
Phys. Rev. B 89, 014401 (2014).


\bibitem{barghathi15} 
 H. Barghathi, F. Hrahsheh, J. A. Hoyos, R. Narayanan and T. Vojta, 
Strong-randomness phenomena in quantum Ashkin-Teller models,
Physica Scripta T165, 014040 (2015).

\bibitem{zhu15}
Q. Zhu, X. Wan, R. Narayanan, J. A. Hoyos and T. Vojta, 
Emerging criticality in the disordered three-color Ashkin-Teller model,
Phys. Rev. B 91, 224201 (2015)

\bibitem{bellafard16}
A. Bellafard and S. Chakravarty
Activated scaling in disorder-rounded first-order quantum phase transitions
Phys. Rev. B 94, 094408 , (2016)


\bibitem{chatelain16} 
 C. Chatelain and D. Voliotis,
Numerical evidence of the double-Griffiths phase of the random quantum Ashkin-Teller chain,
The European Physical Journal B 89, 18 (2016).

\bibitem{ibrahim17} 
A. K. Ibrahim and T. Vojta, 
Monte Carlo simulations of the disordered three-color quantum Ashkin-Teller chain, 
Phys. Rev. B 95, 054403 (2017).





\bibitem{lamas06}
C. A. Lamas, D. C. Cabra, M. D. Grynberg, and G. L. Rossini,
Comparison between disordered quantum spin-1/2 chains,
Phys. Rev. B 74, 224435  (2006).

\bibitem{kokalj}
J. Kokalj, J. Herbrych, A. Zheludev and P. Prelovsek,
Antiferromagnetic order in weakly coupled random spin chains
Phys. Rev. B 91, 155147 (2015)



\bibitem{yu06AF2d}
R. Yu, T. Roscilde and S. Haas,
Quantum disorder and Griffiths singularities in bond-diluted two-dimensional Heisenberg antiferromagnets,
Phys. Rev. B 73, 064406 (2006).

\bibitem{ma14AF2d}
N. Ma, A. W. Sandvik and D.X. Yao
Criticality and Mott glass phase in a disordered two-dimensional quantum spin system
Phys. Rev. B 90 104425 (2014)


\bibitem{liu18AFd2}
L. Liu, H. Shao, Y.C. Lin, W. Guo and A. W. Sandvik,
Random-Singlet Phase in Disordered Two-Dimensional Quantum Magnets,
arxiv 1804.06108.


 \bibitem{zhou09}
 S. Zhou, J. A. Hoyos, V. Dobrosavljevic and E. Miranda, 
 Valence-bond theory of highly disordered quantum antiferromagnets,
Europhys. Lett. 87, 27003 (2009).
  
\bibitem{quito15} 
V. L. Quito, J. A. Hoyos and E. Miranda, 
 Emergent SU(3) symmetry in random spin-1 chains, 
Phys. Rev. Lett. 115, 167201 (2015).
 
 \bibitem{quito16b} 
 V. L. Quito, J. A. Hoyos and E. Miranda, 
Random SU(2)-symmetric spin-S chains,
Phys. Rev. B 94, 064405 (2016).

 \bibitem{quito17a} 
 V. L. Quito, Pedro L. S. Lopes, J. A. Hoyos and E. Miranda 
Highly-symmetric random one-dimensional spin models, 
arXiv:1711.04783.
 
 \bibitem{quito17b} 
 V. L. Quito, Pedro L. S. Lopes, J. A. Hoyos and E. Miranda 
Emergent SU(N) symmetry in disordered SO(N) spin chains,
arXiv:1711.04781.


\bibitem{lajko06} 
P. Lajk\'o,
Renormalization-group investigation of the S=1 random antiferromagnetic Heisenberg Chain, 
 Int. J. Mod. Phys. C 17, 1739 (2006).



\bibitem{lavarelo12}
A. Lavarelo and G. Roux,
Localization of Spinons in Random Majumdar-Ghosh Chains,
Phys. Rev. Lett. 110, 087204 (2013).


\bibitem{strinati17}
M. C. Strinati, D. Rossini, R. Fazio and A. Russomanno,
Resilience of hidden order to symmetry-preserving disorder,
Phys. Rev. B 96, 214206 (2017).




\bibitem{melin06} 
R. M\'elin and F. Igl\'oi, 
Strongly disordered Hubbard model in one dimension: spin and orbital infinite randomness and Griffiths phases, 
Phys. Rev. B 74, 155104 (2006).

\bibitem{lin06} 
 Y.C. Lin, H. Rieger, N. Laflorencie and F. Igl\'oi, 
Strong disorder renormalization group study of S=1/2 Heisenberg antiferromagnet layers/bilayers with bond randomness, site dilution and dimer dilution,
Phys. Rev. B 74, 024427 (2006).






\bibitem{bonesteel07} 
 N.E. Bonesteel and K. Yang, 
Infinite-Randomness Fixed Points for Chains of Non-Abelian Quasiparticles,
Phys. Rev. Lett. 99, 140405 (2007).

\bibitem{fidkowski08a} 
 L. Fidkowski, G. Refael, N. Bonesteel and J. Moore,
c-theorem violation for effective central charge of infinite-randomness fixed points,
Phys. Rev. B 78, 224204 (2008).

\bibitem{fidkowski08b} 
 L. Fidkowski, G. Refael, H.H. Lin and P. Titum,
Permutation Symmetric Critical Phases in Disordered Non-Abelian Anyonic Chains,
Phys. Rev. B 79, 155120 (2009).

\bibitem{refael09} 
 G. Refael, J. E. Moore,
Criticality and entanglement in random quantum systems,
J. Phys. A: Math. Theor. 42 504010 (2009).



 





\bibitem{altman07} 
 E. Altman, Y. Kafri, A. Polkovnikov and G. Refael 
The insulating phases and superfluid-insulator transition of disordered boson chains,
Phys. Rev. Lett. 100, 170402 (2008)

\bibitem{gurarie08}
V. Gurarie, G. Refael and J. T. Chalker,
Excitations of the One Dimensional Bose-Einstein Condensates in a Random Potential,
Phys. Rev. Lett. 101, 170407 (2008).
 
\bibitem{altman09} 
 E. Altman, Y. Kafri, A. Polkovnikov and G. Refael,
Superfluid-insulator transition of disordered bosons in one-dimension,
Phys. Rev. B 81, 174528 (2010).

\bibitem{vosk12superfluid}
R. Vosk and E. Altman,
Superfluid-insulator transition of ultracold bosons in disordered one-dimensional traps,
Phys. Rev. B 85 024531 (2012).
 
\bibitem{iyer13} 
 S. Iyer, D. Pekker and G. Refael, 
Susceptibility at the Superfluid-Insulator Transition for One-Dimensional Disordered Bosons,
Phys. Rev. B 88, 220501 (2013).

\bibitem{refael13}
 G. Refael and E. Altman, 
Strong disorder renormalization group primer and the superfluid-insulator transition, 
Comptes Rendus Physique 14 , 725 (2013).

\bibitem{hrahsheh12}
F. Hrahsheh and T. Vojta,
Disordered bosons in one dimension: from weak to strong randomness criticality,
Phys. Rev. Lett. 109, 265303 (2012)


\bibitem{doggen17}
E. V. H. Doggen, G. Lemarié, S. Capponi and N. Laflorencie,
Weak Versus Strong Disorder Superfluid-Bose Glass Transition in One Dimension,
Phys. Rev. B 96, 180202 (2017).




\bibitem{maestro08}
A. Del Maestro, B. Rosenow, M. Muller and S. Sachdev,
Infinite Randomness Fixed Point of the Superconductor-Metal Quantum Phase Transition,
Phys. Rev. Lett. 101, 035701 (2008).


\bibitem{maestro10}
A. Del Maestro, B. Rosenow, J. A. Hoyos and T. Vojta,
Dynamical Conductivity at the Dirty Superconductor-Metal Quantum Phase Transition,
Phys. Rev. Lett. 105, 145702 (2010).

\bibitem{nozadze14}
D. Nozadze and T. Vojta, 
Numerical method for disordered quantum phase transitions in the large-N limit,
 Physica Status Solidi 251 675 (2014).
 
\bibitem{ibrahim18}
A. K. Ibrahim and T. Vojta,
Monte Carlo simulations of a disordered superconductor-metal quantum phase transition,
arXiv:1808.00094
 




\bibitem{ramirez14}
G. Ramirez, J. Rodriguez-Laguna and G. Sierra,
From conformal to volume-law for the entanglement entropy in exponentially deformed critical spin 1/2 chains,
 J. Stat. Mech.  P10004 (2014).

\bibitem{ramirez15}
G. Ramirez, J. Rodriguez-Lagunaand G. Sierra, 
Entanglement over the rainbow,
J. Stat. Mech. P06002 (2015).

\bibitem{rodriguez16}
J. Rodriguez-Laguna, S. N. Santalla, G. Ramirez and G. Sierra,  
Entanglement in correlated random spin chains, RNA folding and kinetic roughening,
New J. Phys. 18 073025 (2016).


\bibitem{rodriguez17}
J. Rodriguez-Laguna, J. Dubail, G. Ramirez, P. Calabrese, G. Sierra, 
More on the rainbow chain: entanglement, space-time geometry and thermal states,
J. Phys. A: Math. Theor. 50 164001 (2017).

\bibitem{ramirez18}
G. Ramirez,
Quantum Entanglement In Inhomogeneous 1D Systems,
AIP Conference Proceedings 1950, 030007 (2018).

\bibitem{alba18}
V. Alba, S. N. Santalla, P. Ruggiero, J. Rodriguez-Laguna, P. Calabrese and G. Sierra,
Unusual area-law violation in random inhomogeneous systems,
arxiv : 1807.04179




\bibitem{calabrese09} 
P. Calabrese, J. Cardy and B. Doyon (Eds. of the special issue)
Entanglement entropy in extended quantum systems,
 J. Phys. A 42 500301 (2009).

\bibitem{amico08} 
L. Amico, R. Fazio, A. Osterloh, and V. Vedral, 
Entanglement in many-body systems
Rev. Mod. Phys. 80, 517 (2008).

\bibitem{laflorencie16} 
 N. Laflorencie, 
Quantum entanglement in condensed matter systems,
Phys. Rep. 643, 1 (2016).

\bibitem{holzhey94} 
C. Holzhey, F. Larsen, and F. Wilczek, 
Geometric and Renormalized Entropy in Conformal Field Theory
Nucl. Phys. B 424, 443 (1994).

\bibitem{vidal03} 
G. Vidal, J. I. Latorre, E. Rico, and A. Kitaev, 
Entanglement in Quantum Critical Phenomena,
Phys. Rev. Lett. 90, 227902 (2003).

\bibitem{calabrese04} 
P. Calabrese and J. Cardy, 
Entanglement Entropy and Quantum Field Theory,
J. Stat. Mech. (2004) P06002.

\bibitem{calabrese08}
P. Calabrese and A. Lefevre, 
Entanglement spectrum in one-dimensional systems,
Phys. Rev. A 78, 32329 (2008).

\bibitem{vasseur17}
R. Vasseur, A. Roshani, S. Haas and H. Saleur, 
Healing of Defects in Random Antiferromagnetic Spin Chains, 
EPL 119 50004 (2017).

\bibitem{vasseur15multif}
R. Vasseur and J. E. Moore,
Multifractal Orthogonality Catastrophe in 1D Random Quantum Critical Points,
Phys. Rev. B 92, 054203 (2015)

 
\bibitem{hoyos07} 
 J. A. Hoyos, A. P. Vieira, N. Laflorencie, E. Miranda, 
Correlation amplitude and entanglement entropy in random spin chains,
Phys. Rev. B 76, 174425 (2007).
 
\bibitem{refael07} 
 G. Refael and J. E. Moore, 
Entanglement entropy of the random spin-1 Heisenberg chain,
Phys. Rev. B 76, 024419 (2007).

\bibitem{saguia07} 
 A. Saguia, M. S. Sarandy, B. Boechat and M. A. Continentino,
Entanglement Entropy in Random Quantum Spin-S Chains,
Phys. Rev. A 75, 052329 (2007).

\bibitem{saguia10} 
 A. Saguia and M. S. Sarandy, 
Nonadditive entropy for random quantum spin-S chains,
Phys. Lett. A 374, 3384 (2010).

\bibitem{igloi08} 
 F. Igl\'oi and R. Juh\'asz, 
Exact relationship between the entanglement entropies of XY and quantum Ising chains,
Europhys. Lett. 81, 57003 (2008).

 \bibitem{igloi07} 
F. Igl\'oi, Y.C. Lin, H. Rieger and C. Monthus, 
Finite-size scaling of pseudo-critical point distributions in the random transverse-field Ising chain, 
Phys. Rev. B 76, 064421 (2007).


\bibitem{santachiara06} 
 R. Santachiara, 
Increasing of entanglement entropy from pure to random quantum critical chains,
J. Stat. Mech.  L06002 (2006).

\bibitem{hoyos11} 
 J. A. Hoyos, N. Laflorencie, A. P. Vieira and T. Vojta, 
Protecting clean critical points by local disorder correlations,
Europhys. Lett. 93, 30004 (2011).

\bibitem{getelina16} 
 J. C. Getelina, F. C. Alcaraz, J. A. Hoyos, 
Entanglement properties of correlated random spin chains and similarities with conformal invariant systems,
Phys. Rev. B 93, 045136 (2016).

\bibitem{juhasz17a} 
 R. Juh\'asz, I. A. Kov\'acs, G. Ro\'osz and F. Igl\'oi, 
Entanglement between random and clean quantum spin chains,
J. Phys. A: Math. Theor. 50 , 324003 (2017).

\bibitem{juhasz17b} 
 R. Juh\'asz, 
Entanglement across extended random defects in the XX spin chain,
J. Stat. Mech.  083107 (2017).

\bibitem{juhasz18}
R. Juh\'asz, J. M. Oberreuter and Z. Zimbor\'as,
Entanglement Entropy of Disordered Quantum Wire Junctions,
arXiv:1808.02576

\bibitem{vieira05} 
 A. P. Vieira, 
 Aperiodic quantum XXZ chains: Renormalization-group results,
Phys. Rev. B 71, 134408 (2005).
 
\bibitem{filho12}  
 F. J. O. Filho, M. S. Faria and A. P. Vieira,
Strong disorder renormalization group study of aperiodic quantum Ising chains,,
 J. Stat. Mech.  P03007 (2012).

\bibitem{casa14}
 H. L. Casa Grande, N. Laflorencie, F. Alet and A. P. Vieira, 
Analytical and numerical studies of disordered spin-1 Heisenberg chains with aperiodic couplings,
Phys. Rev. B 89, 134408 (2014).

\bibitem{vieira18}
A. P. Vieira and J. A. Hoyos,
Localization and emergent dimerization in aperiodic quantum spin chains,
arXiv:1804.05108 

 \bibitem{igloi07b} 
F. Igl\'oi, R. Juh\'asz and Z. Zimbor\'as, 
Entanglement entropy of aperiodic quantum spin chains, 
Europhys. Lett. 79, 37001 (2007).
 
 \bibitem{juhasz07b} 
 R. Juh\'asz and Z. Zimbor\'as,  
Entanglement entropy in aperiodic singlet phases,
J. Stat. Mech.  P04004 (2007).

\bibitem{hoyos06channels}
J. A. Hoyos and G. Rigolin, 
Quantum channels in random spin chains
Phys. Rev. A 74, 062324 (2006).

\bibitem{getelina17}
J. C. Getelina, T. R. de Oliveira, J. A. Hoyos
Violation of the Bell inequality in quantum critical random spin-1/2 chains
 arXiv:1711.10005


\bibitem{hide11}
J. Hide,
Concurrence in disordered systems
J. Phys. A 45 115302 (2012).
  
 \bibitem{fagotti11} 
M. Fagotti, P. Calabrese and J. E. Moore, 
Entanglement spectrum of random-singlet quantum critical points, 
Phys. Rev. B83, 045110 (2011).


\bibitem{ramirez14ent}
G. Ramirez, J. Rodriguez-Laguna and G. Sierra,
Entanglement in low-energy states of the random-hopping model,
J. Stat. Mech.  P07003 (2014).



\bibitem{tran09}
H. Tran and N. E. Bonesteel
Valence bond entanglement and fluctuations in random singlet phases
Phys. Rev. B 84 144420 (2011)

 
 \bibitem{devakul17} 
 T. Devakul, S. N. Majumdar and D. A. Huse, 
Probability distribution of the entanglement across a cut at an infinite-randomness fixed point,
Phys. Rev. B 95, 104204 (2017).
 
 \bibitem{torlai18} 
 G. Torlai, K. D. McAlpine, G. De Chiara
Schmidt gap in random spin chains,
 arXiv:1805.07404.

 \bibitem{ruggiero16} 
 P. Ruggiero, V. Alba and P. Calabrese, 
The entanglement negativity in random spin chains,
Phys. Rev. B 94, 035152 (2016).

\bibitem{alet07} 
F. Alet, S. Capponi, N. Laflorencie and M. Mambrini, 
Valence Bond Entanglement Entropy, 
Phys. Rev. Lett. 99, 117204 (2007).
 
\bibitem{lin10} 
 Y.C. Lin and A. W. Sandvik, 
Definitions of entanglement entropy of spin systems in the valence-bond basis,
Phys. Rev. B 82, 224414 (2010).

\bibitem{shu16}
 Y.-R. Shu, D.-X. Yao, C.-W. Ke, Y.-Ch. Lin and A. W. Sandvik, 
Properties of the random-singlet phase: from the disordered Heisenberg chain to an amorphous valence-bond solid,
Phys. Rev. B 94, 174442 (2016).

\bibitem{kovacs12a} 
 I. A. Kov\'acs and F. Igl\'oi, 
Universal logarithmic terms in the entanglement entropy of 2d, 3d and 4d random transverse-field Ising models,
EPL 97, 67009 (2012).

\bibitem{senthil96} 
T. Senthil and S. Sachdev, 
Higher Dimensional Realizations of Activated Dynamic Scaling at Random Quantum Transitions,
Phys. Rev. Lett. 77, 5292 (1996).

\bibitem{vojta07}
T. Vojta and J. A. Hoyos, Quantum Phase Transitions on Percolating Lattices,
arXiv:0707.0658

\bibitem{kovacs12b} 
 I. A. Kov\'acs, F. Igl\'oi and J. Cardy, 
Corner contribution to percolation cluster numbers,
Phys. Rev. B 86, 214203 (2012).

\bibitem{cardy88} 
J. Cardy and I. Peschel, 
Finite-size dependence of the free energy in two-dimensional critical systems
Nucl. Phys. B, 300 , 377 (1988).

\bibitem{kovacs14} 
 I. A. Kov\'acs and F. Igl\'oi, 
Corner contribution to percolation cluster numbers in three dimensions,
Phys. Rev. B. 89 174202 (2014).



\bibitem{cirac09}
J. I. Cirac and F. Verstraete, 
Renormalization and tensor product states in spin chains and lattices,
J. Phys. A: Math. Theor. 42, 504004 (2009).

\bibitem{gittsovich10}
O. Gittsovich, R. Hubener, E. Rico and H.J. Briegel,
Local renormalization method for random systems
New J. Phys. 12, 025020 (2010).

\bibitem{goldsborough14}
A. M. Goldsborough and R. A. Romer, 
Self-assembling tensor networks and holography in disordered spin chains,
Phys. Rev. B 89, 214203 (2014).



\bibitem{goldsborough17}
A. M. Goldsborough and G. Evenbly, 
Entanglement renormalization for disordered systems,
Phys. Rev. B 96, 155136 (2017).

\bibitem{chatelain18}
C. Chatelain,
Quantifying and improving the accuracy of the Matrix Product Operator Renormalization Group of random spin chains,
arxiv: 1807.08984



\bibitem{hyatt17}
K. Hyatt, J. R. Garrison and B. Bauer, 
Extracting entanglement geometry from quantum states,
Phys. Rev. Lett. 119, 140502 (2017).



 \bibitem{nandkishore15}
R. Nandkishore and D. A. Huse, 
Many body localization and thermalization in quantum statistical mechanics,
Ann. Review of Cond. Mat. Phys. 6, 15 (2015).

 \bibitem{altman15mblreview}
 E. Altman and R. Vosk, 
Universal dynamics and renormalization in many body localized systems,
Ann. Review of Cond. Mat. Phys. 6, 383 (2015).

\bibitem{parameswaran17}
S. A. Parameswaran, A. C. Potter and R. Vasseur, 
Eigenstate phase transitions and the emergence of universal dynamics in highly excited states,
Annalen der Physik  529, 1600302 (2017).

 \bibitem{alet18}
F. Alet and N. Laflorencie, 
Many-body localization: an introduction and selected topics,
Comptes Rendus Physique (2018).

\bibitem{abanin18}
D. A. Abanin, E. Altman, I. Bloch and M. Serbyn,
Ergodicity, Entanglement and Many-Body Localization,
 arxiv 1804.11065

\bibitem{imbrie17}
J. Z. Imbrie, V. Ros, A. Scardicchio, 
Review: Local Integrals of Motion in Many-Body Localized systems,
 Annalen der Physik 529, 1600278 (2017).

\bibitem{rademaker17}
L. Rademaker, M. Ortuno and A.M. Somoza, 
 Many-body localization from the perspective of Integrals of Motion,
 Annalen der Physik 529, 1600322 (2017).



\bibitem{pekker14rsrgx}
D. Pekker, G. Refael, E. Altman, E. Demler and V. Oganesyan, 
Hilbert-Glass Transition: New Universality of Temperature-Tuned Many-Body Dynamical Quantum Criticality,
Phys. Rev. X 4, 011052 (2014).

 \bibitem{you16}
 Y.Z. You, X.L. Qi and C. Xu, 
Entanglement Holographic Mapping of Many-Body Localized System by Spectrum Bifurcation Renormalization Group,
Phys. Rev. B 93, 104205 (2016).

\bibitem{monthus18rsrgxmaj}
C. Monthus, 
Strong Disorder Real-Space Renormalization for the Many-Body-Localized phase of random Majorana models,
J. Phys. A: Math. Theor. 51 115304 (2018).


 
 
\bibitem{huang14} 
Y. Huang, Joel E. Moore, 
Excited-state entanglement and thermal mutual information in random spin chains, 
Phys. Rev B90, 220202 (2014).



 \bibitem{pouranvari15}
M. Pouranvari and K. Yang, 
Entanglement spectrum and entangled modes of highly excited states in random XX spin chains,
Phys. Rev. B 92, 245134 (2015).


\bibitem{agarwal15} 
 K. Agarwal, E. Demler and I. Martin, 
$1/f^{\alpha}$ noise and generalized diffusion in random Heisenberg spin systems,
Phys. Rev. B 92, 184203 (2015).

\bibitem{vasseur16particlehole}
    R. Vasseur, A. J. Friedman, S. A. Parameswaran and A. C. Potter, 
 Particle-hole symmetry, many-body localization, and topological edge modes,
  Phys. Rev. B 93, 134207 (2016).


\bibitem{slage16}
K. Slagle, Y. Z. You, and C. Xu, 
Disordered XYZ spin chain simulations using the spectrum bifurcation renormalization group,
Phys. Rev. B 94, 014205 (2016).

\bibitem{friedman17}
A. J. Friedman, R. Vasseur, A. C. Potter, S. A. Parameswaran,
Localization-protected order in spin chains with non-Abelian discrete symmetries,
arxiv 1706.00022

 \bibitem{vasseur15hot}
R. Vasseur, A. C. Potter and S.A. Parameswaran, 
Quantum Criticality of Hot Random Spin Chains,
Phys. Rev. Lett. 114, 217201 (2015).

\bibitem{kang17} 
B. Kang, A. C. Potter and R. Vasseur,
 Universal crossover from ground state to excited-state quantum criticality, 
Phys. Rev. B 95, 024205 (2017).


\bibitem{monthus16emergent}
C. Monthus,
Many-Body Localization : construction of the emergent local conserved operators via block real-space renormalization,
  J. Stat. Mech.  033101 (2016)


\bibitem{monthus17mblcayley}
C. Monthus, 
Random Transverse Field Spin-Glass Model on the Cayley tree : phase transition between the two Many-Body-Localized Phases,
J. Stat. Mech. 123304 (2017).




 \bibitem{vosk13}
R. Vosk and E. Altman, 
Many-body localization in one dimension as a dynamical renormalization group fixed point,
Phys. Rev. Lett. 110, 067204 (2013).

 \bibitem{vosk14}
R. Vosk and E. Altman, 
Dynamical Quantum Phase Transitions in Random Spin Chains,
Phys. Rev. Lett. 112, 217204 (2014).

\bibitem{bukov15}
M. Bukov, L. D'Alessio and A. Polkovnikov, 
Universal High-Frequency Behavior of Periodically Driven Systems: from Dynamical Stabilization to Floquet Engineering,
Advances in Physics, Vol. 64, No. 2, 139 (2015).

\bibitem{monthus17rsrgt}
C. Monthus,
Strong Disorder Renormalization for the dynamics of Many-Body-Localized systems : iterative elimination of the fastest degree of freedom via the Floquet expansion,
J. Phys. A: Math. Theor. 51 275302 (2018).

 \bibitem{huang17}
Y. Huang, 
Entanglement dynamics in critical random quantum Ising chain with perturbations,
Annals of Physics 380, 224 (2017).



\bibitem{heyl15}
M. Heyl and M. Vojta, 
Nonequilibrium dynamical renormalization group: Dynamical crossover from weak to infinite randomness in the transverse-field Ising chain, 
Phys. Rev. B 92, 104401 (2015).

\bibitem{hauke15}
P. Hauke and M. Heyl,
Many-body localization and quantum ergodicity in disordered long-range Ising models,
 Phys. Rev. B 92, 134204 (2015).




\bibitem{chiara06}
G. De Chiara, S. Montangero, P. Calabrese, R. Fazio,
Entanglement Entropy dynamics in Heisenberg chains,
 J. Stat. Mech., L03001 (2006).

\bibitem{igloi12}
F. Igl\'oi, Z. Szatm\'ari, and Y.-C. Lin,
Entanglement entropy dynamics of disordered quantum spin chains
Phys. Rev. B 85, 094417 (2012).

\bibitem{levine12}
G. C. Levine, M. J. Bantegui, J. A. Burg, 
Full counting statistics in a disordered free fermion system,
Phys. Rev. B 86, 174202 (2012).

\bibitem{bardarson12}
J. H. Bardarson, F. Pollmann, and J. E. Moore,
Unbounded Growth of Entanglement in Models of Many-Body Localization,
Phys. Rev. Lett. 109, 017202 (2012).

\bibitem{zhao16}
Y. Zhao, F. Andraschko, J. Sirker, 
Entanglement entropy of disordered quantum chains following a global quench,
Phys. Rev. B 93, 205146 (2016).



\bibitem{roosz17} 
 G. Ro\'osz, Y.C. Lin and F. Igl\'oi, 
Critical quench dynamics of random quantum spin chains: Ultra-slow relaxation from initial order and delayed ordering from initial disorder,
New J. Phys. 19, 023055 (2017).


 \bibitem{herbrych13}
J. Herbrych, J. Kokalj, and P. Prelovsek,
Local Spin Relaxation within the Random Heisenberg Chain,
Phys. Rev. Lett. 111 147203 (2013).


\bibitem{shu17} 
Y.R. Shu, M. Dupont, D.X. Yao, S. Capponi and A. W. Sandvik 
Dynamical properties of the S=1/2 random Heisenberg chain, 
Phys. Rev. B 97, 104424 (2018).


\bibitem{igloi13} 
 F. Igl\'oi, G. Ro\'osz and Y.C. Lin, 
Nonequilibrium quench dynamics in quantum quasicrystals,
New J. Phys. 15, 023036 (2013).
 
\bibitem{roosz14} 
G. Ro\'osz, U. Divakaran, H. Rieger and F. Igl\'oi,
Non-equilibrium quantum relaxation across a localization-delocalization transition
 Phys. Rev. B 90, 184202 (2014).

\bibitem{divakaran18}
U. Divakaran, Sudden quenches in quasiperiodic Ising model,
arXiv:1805.07977

\bibitem{mason17}
P. Mason, A. M. Zagoskin and J. J. Betouras,
Time-dependent Real-space Renormalization-Group Approach:  application to an adiabatic random quantum Ising model,
arxiv 1708.05948




\bibitem{monthus10mblaoki}
C. Monthus and T. Garel, 
Many-body localization transition in a lattice model of interacting fermions: statistics of renormalized hoppings in configuration space,
Phys. Rev. B 81, 134202 (2010).


\bibitem{vosk15mbltransition}
R. Vosk, D.A. Huse, and E. Altman, 
Theory of the many-body localization transition in one dimensional systems,
Phys. Rev. X 5, 031032 (2015).

\bibitem{potter15mbltransition}
A. C. Potter, R. Vasseur and S.A. Parameswaran, 
Universal properties of many-body delocalization transitions,
Phys. Rev. X 5, 031033 (2015).

\bibitem{dumitrescu17}
P. T. Dumitrescu, R. Vasseur, A. C. Potter, 
Scaling Theory of Entanglement at the Many-Body Localization Transition,
Phys. Rev. Lett. 119, 110604 (2017).

\bibitem{zhang16}
L. Zhang, B. Zhao, T. Devakul and D. A. Huse, 
Many-body localization phase transition: A simplified strong-randomness approximate renormalization group,
Phys. Rev. B 93, 224201 (2016).

\bibitem{goremykina}
A. Goremykina, R. Vasseur and M. Serbyn,
Analytically solvable renormalization group for the many-body localization transition,
arxiv: 1807.04285


\bibitem{kehrein06}
S. Kehrein, The flow equation approach to many-particle systems, Springer-Verlag Berlin (2006). 

\bibitem{rademaker16}
L. Rademaker and M. Ortuno, 
Explicit Local Integrals of Motion for the Many-Body Localized State,
Phys. Rev. Lett. 116, 010404 (2016).


\bibitem{monthus16}
C. Monthus, 
Flow towards diagonalization for Many-Body-Localization models : adaptation of the Toda matrix differential flow to random quantum spin chains,
J. Phys. A: Math. Theor. 49 305002 (2016).

\bibitem{pekker17wegner}
D. Pekker, B. K. Clark, V. Oganesyan and G. Refael, 
Fixed points of Wegner-Wilson flows and many-body localization,
Phys. Rev. Lett. 119, 075701 (2017).

\bibitem{savitz17}
S. Savitz and G. Refael, 
Stable Unitary Integrators for the Numerical Implementation of Continuous Unitary Transformations,
Phys. Rev. B 96, 115129 (2017).

\bibitem{thomson18}
S. J. Thomson and M. Schiro, 
Time Evolution of Many-Body Localized Systems with the Flow Equation Approach,
Phys. Rev. B 97, 060201 (2018).







\bibitem{moessner17}
R. Moessner and S. L. Sondhi, 
Equilibration and Order in Quantum Floquet Matter,
Nature Physics 13,  424 (2017)

\bibitem{monthus17floquet}
C. Monthus,
Periodically driven random quantum spin chains : Real-Space Renormalization for Floquet localized phases,
 J. Stat. Mech. (2017) 073301

 
\bibitem{vasseur18}
W. Berdanier, M. Kolodrubetz, S. A. Parameswaran and R. Vasseur, 
Floquet Quantum Criticality,
arXiv:1803.00019

\bibitem{berdanier}
W. Berdanier, M. Kolodrubetz, S. A. Parameswaran and R. Vasseur,
Strong-Disorder Renormalization Group for Periodically Driven Systems,
arxiv : 1807.09767







\bibitem{weiss99}
U. Weiss,  
Quantum dissipative systems,
 Singapore: World Scientific, 2nd edition (1999).

\bibitem{breuer02}
 H. P. Breuer and F. Petruccione, 
The theory of open quantum systems 
Oxford (2002).


\bibitem{leggett87}
Leggett, A., S. Chakravarty, A. Dorsey, M. Fisher, A. Garg, and W. Zwerger, 
Dynamics of the dissipative two-state system,
 Rev. Mod. Phys. 59, 1 (1987).


\bibitem{schehr06}
G. Schehr and H. Rieger, 
Strong Randomness Fixed Point in the Dissipative Random Transverse Field Ising Model,
Phys. Rev. Lett. 96, 227201 (2006).

\bibitem{schehr08}
G. Schehr and H. Rieger,
Finite temperature behavior of strongly disordered quantum magnets coupled to a dissipative bath
 J. Stat. Mech. P04012  (2008).

\bibitem{hoyos08dissipation}
J. A. Hoyos and T. Vojta, 
Dissipation effects in percolating quantum Ising magnets,
Physica B 403, 1245 (2008).

\bibitem{hoyos08smeared}
J. A. Hoyos and T. Vojta, 
Theory of Smeared Quantum Phase Transitions,
Phys. Rev. Lett. 100, 240601 (2008).

\bibitem{vojta10}
T. Vojta and J. A. Hoyos, 
Smeared quantum phase transition in the dissipative random quantum Ising model,
Physica E 42, 383 (2010).


\bibitem{hoyos12}
J. A. Hoyos and T. Vojta, 
Dissipation effects in random transverse-field Ising chains,
Phys. Rev. B 85, 174403 (2012).

\bibitem{ali13}
M. Al-Ali and T. Vojta,
Monte-Carlo simulations of the dissipative random transverse-field Ising chain
arxiv 1307.7166


\bibitem{hoyos07LG}
J. A. Hoyos, C. Kotabage and T. Vojta, 
Effects of dissipation on a quantum critical point with disorder,
Phys. Rev. Lett. 99, 230601 (2007).

\bibitem{vojta09LG}
T. Vojta, C. Kotabage and J. A. Hoyos, 
Infinite-randomness quantum critical points induced by dissipation,
Phys. Rev. B 79, 024401 (2009).

\bibitem{vojta11LG}
T. Vojta, J. A. Hoyos, P. Mohan and R. Narayanan, 
Influence of superohmic dissipation on a disordered quantum critical point,
 J. Phys.: Condens. Matter 23, 094206 (2011).


\bibitem{monthus17step}
C. Monthus,
Boundary-driven Lindblad dynamics of random quantum spin chains : strong disorder approach for the relaxation, the steady state and the current,
 J. Stat. Mech. 043303 (2017).

\bibitem{roeck17}
W. De Roeck, A. Dhar, F. Huveneers and M. Schutz,
Step Density Profiles in Localized Chains
J.  Stat. Phys. 167 , 1143 (2017).

\bibitem{monthus17dephasing}
C. Monthus, 
Dissipative random quantum spin chain with boundary-driving and bulk-dephasing: magnetization and current statistics in the Non-Equilibrium-Steady-State,
J. Stat. Mech. 043302 (2017).



\bibitem{chatelain17}
C. Chatelain,
Diverging conductance at the contact between random and pure quantum XX spin chains,
 J. Stat. Mech.  113301 (2017).




\bibitem{evers08}
F. Evers and A. D. Mirlin,
Anderson transitions
Rev. Mod. Phys. 80 1355 (2008).

\bibitem{mard14}
H. J. Mard, J. A. Hoyos, E. Miranda and V. Dobrosavljevic,  
Strong-disorder renormalization-group study of the one-dimensional tight-binding model,
Phys. Rev. B 90,125141 (2014).

\bibitem{mard17}
H. J. Mard, J. A. Hoyos, E. Miranda and V. Dobrosavljevic, 
Strong-disorder approach for the Anderson localization transition,
Phys. Rev. B 96, 045143 (2017).

\bibitem{aoki80}
H. Aoki,
Real-space renormalisation-group theory for Anderson localisation: decimation method for electron systems,
 J. Phys. C: Solid State Phys. 13 3369 (1980).

\bibitem{aoki82}
H. Aoki, 
Decimation method of real-space renormalization for electron systems with application to random systems,
Physica A 114, 538 (1982).

\bibitem{monthus09aoki}
C. Monthus and T. Garel, 
Statistics of renormalized on-site energies and renormalized hoppings for Anderson localization models in dimensions d=2 and d=3,
Phys. Rev. B 80, 024203 (2009).

\bibitem{tarquini17}
E. Tarquini, G. Biroli and M. Tarzia, 
Critical properties of the Anderson localization transition and the high dimensional limit,
Phys. Rev. B 95, 094204 (2017).

\bibitem{johri14}
S. Johri and R. N. Bhatt,
Large Disorder Renormalization Group Study of the Anderson Model of Localization
Phys. Rev. B 90, 060205(R) (2014).

\bibitem{bhatt12}
R.N. Bhatt and S. Johri,
Rare Fluctuation Effects in the Anderson Model of Localization,
Int. J. Mod. Phys. Conf. Ser. 11 79 (2012).


\bibitem{monthus10cascade}
C. Monthus and T. Garel, 
Random cascade models of multifractality : real-space renormalization and travelling-waves,
J. Stat. Mech. P06014 (2010).

\bibitem{monthus11dysonanderson}
C. Monthus and T. Garel,
A critical Dyson hierarchical model for the Anderson localization transition
 J. Stat. Mech. P05005 (2011).


\bibitem{quito16}
V. L. Quito, P. Titum, D. Pekker and G. Refael, 
Localization transition in one dimension using Wegner flow equations
Phys. Rev. B 94, 104202 (2016).


\bibitem{harris74} 
T.E. Harris, 
Contact Interactions on a Lattice,
Ann. Prob.,  2, 969 (1974).

\bibitem{liggett05} 
T.M. Liggett, 
 Stochastic interacting systems:  contact, voter, and exclusion processes,
 Berlin Springer  (2005).

\bibitem{vojta08mc}
T. Vojta, A. Farquhar and J. Mast,
Infinite-randomness critical point in the two-dimensional disordered contact process,
Phys. Rev. E 79 011111 (2009)

 
\bibitem{vojta09} 
T. Vojta, A. Farquhar and J. Mast, 
Infinite-randomness critical point in the two- dimensional disordered contact process, 
Phys. Rev. E 79, 011111 (2009)
 
\bibitem{vojta12} 
 T. Vojta, 
Monte-Carlo simulations of the clean and disordered contact process in three dimensions,
Phys. Rev. E 86, 051137 (2012)

 
\bibitem{hoyos08} 
 J. A. Hoyos, 
Weakly disordered absorbing-state phase transitions,
Phys. Rev. E 78, 032101 (2008).

\bibitem{juhasz13dyn}
R. Juh\'asz,
Distribution of dynamical quantities in the contact process, random walks, and quantum spin chains in random environments,
Phys. Rev. E 89, 032108 (2014)

\bibitem{juhasz13a}
 R. Juh\'asz, 
Disordered contact process with asymmetrics preading,
Phys. Rev. E 87, 022133 (2013).

\bibitem{vojta14five}
T. Vojta, J. Igo and  J. A. Hoyos,
Rare regions and Griffiths singularities at a clean critical point: The five-dimensional disordered contact process
Phys. Rev. E 90, 012139 (2014).


\bibitem{vojta14rareharris}
T. Vojta and J. A. Hoyos,
Criticality and quenched disorder: rare regions vs. Harris criterion
 Phys. Rev. Lett. 112, 075702 (2014)


\bibitem{ibrahim14}
A. K. Ibrahim, H. Barghathi and T. Vojta,
Enhanced rare region effects in the contact process with long-range correlated disorder
Phys. Rev. E 90, 042132 (2014).

\bibitem{barghathi14}
H. Barghathi, D. Nozadze and T. Vojta,
Contact process on generalized Fibonacci chains: infinite-modulation criticality and double-log periodic oscillations,
Phys. Rev. E 89, 012112 (2014)


\bibitem{munoz10complex}
M. A. Munoz, R. Juh\'asz, C. Castellano and G. \'Odor,
Griffiths Phases on Complex Networks
Phys. Rev. Lett. 105 128701 (2010).

\bibitem{vojta12and15}
H. Barghathi and T. Vojta,
Random fields at a nonequilibrium phase transition
Phys. Rev. Lett. 109, 170603 (2012) ; 
H. Barghathi and T. Vojta,
Random field disorder at an absorbing state transition in one and two dimensions,
Phys. Rev. E 93, 022120 (2016).



\bibitem{juhasz15} 
 R. Juh\'asz, I. A. Kov\'acs and F. Igl\'oi, 
Long-range epidemic spreading in a random environment,
Phys. Rev. E 91, 032815 (2015).

\bibitem{juhasz13b} 
 R. Juh\'asz, I. A. Kov\'acs, 
Infinite randomness critical behavior of the contact process on networks with long-range connections,
J. Stat. Mech. P06003 (2013).

\bibitem{vojta15} 
 T. Vojta and J. A. Hoyos, 
 Infinite-noise criticality: Nonequilibrium phase transitions in fluctuating environments,
Europhys. Lett. 112, 30002 (2015).

\bibitem{vazquez11}
F. Vazquez, J. A. Bonachela, C. Lopez and M. A. Munoz,
Temporal Griffiths Phases,
Phys. Rev. Lett. 106 235702 (2011)


\bibitem{barghathi16} 
H. Barghathi, J. A. Hoyos and T. Vojta, 
Contact process with temporal disorder, 
Phys. Rev. E 94, 022111 (2016).

\bibitem{fiore18}
C. E. Fiore, M. M. de Oliveira, J. A. Hoyos,
Temporal disorder in discontinuous non-equilibrium phase transitions: general results,
arXiv:1806.10421

\bibitem{vojta16spatiotemp}
T. Vojta and R. Dickman
Spatio-temporal generalization of the Harris criterion and its application to diffusive disorder
Phys. Rev. E 93, 032143 (2016).




\bibitem{ledoussal09}
P. Le Doussal,
The Sinai model in the presence of dilute absorbers,
 J. Stat. Mech. P07032 (2009).

\bibitem{juhasz12RWLR}
R. Juh\'asz,
Competition between quenched disorder and long-range connections: A numerical study of diffusion,
Phys. Rev. E 85, 011118 (2012).


\bibitem{juhasz08RWstrip}
R. Juh\'asz, 
Random walks in a random environment on a strip: a renormalization group approach,
 J. Phys. A: Math. Theor. 41 315001 (2008).

\bibitem{juhasz10channel}
R. Juh\'asz and  F. Igl\'oi,
Anomalous diffusion in disordered multi-channel systems,
J. Stat. Mech. P03012 (2010).



\bibitem{juhasz12RWnetwork}
R. Juh\'asz, 
The effect of asymmetric disorder on the diffusion in arbitrary networks,
Europhys. Lett. 98  30001 (2012).



\bibitem{monthus10RWaffine}
C. Monthus and T. Garel,
Random walk in two-dimensional self-affine random potentials: Strong-disorder renormalization approach,
Phys. Rev. E 81, 011138 (2010)


\bibitem{monthus08broad}
C. Monthus and T. Garel,
Non equilibrium dynamics of disordered systems : understanding the broad continuum of relevant time scales via a strong-disorder RG in configuration space,
  J. Phys. A: Math. Theor. 41 , 255002 (2008).

\bibitem{monthus08flow}
C. Monthus and T. Garel,
Non-equilibrium dynamics of finite-dimensional disordered systems : RG flow towards an "infinite disorder" fixed point at large times,
 J. Stat. Mech. P07002 (2008).

\bibitem{monthus08valley}
C. Monthus and T. Garel, 
Equilibrium of disordered systems : constructing the appropriate valleys in each sample via strong disorder renormalization in configuration space,
J. Phys. A: Math. Theor. 41, 375005 (2008).

\bibitem{monthus08interface}
C. Monthus and T. Garel, 
Driven interfaces in random media at finite temperature : is there an anomalous zero-velocity phase at small external force ?
Phys. Rev. E 78, 041133 (2008).

\bibitem{Bo17}
S. Bo and A. Celani, 
Multiple-scale stochastic processes: Decimation, averaging and beyond,
Physics Reports 670, 1 (2017).

\bibitem{monthus10first}
C. Monthus and T. Garel,
Statistics of first-passage times in disordered systems using backward master equations and their exact renormalization rules,
 J. Phys. A: Math. Theor. 43 095001 (2010).

\bibitem{monthus13zerotemperature}
C. Monthus and T. Garel,
Dynamics of Ising models near zero temperature : Real Space Renormalization Approach,
J. Stat. Mech. P02037 (2013).

\bibitem{monthus13cayley}
C. Monthus and T. Garel, 
Dynamical barriers for the random ferromagnetic Ising model on the Cayley tree : traveling-wave solution of the real space renormalization flow,
J. Stat. Mech.  P05012 (2013).

\bibitem{monthus13dyson}
C. Monthus and T. Garel,
Dynamical Barriers in the Dyson Hierarchical model via Real Space Renormalization,
 J. Stat. Mech. P02023 (2013).

\bibitem{monthus14LRSG}
C. Monthus,
Low-temperature dynamics of Long-Ranged Spin-Glasses : full hierarchy of relaxation times via real-space renormalization,
 J. Stat. Mech. P08009 (2014).

\bibitem{monthus16dysonSG}
C. Monthus,
Real-space renormalization for the finite temperature statics and dynamics of the Dyson Long-Ranged Ferromagnetic and Spin-Glass models,
 J. Stat. Mech. 043302 (2016).




\bibitem{dyson53}
F. J. Dyson, 
The Dynamics of a Disordered Linear Chain,
Phys. Rev. 92, 1331 (1953).

\bibitem{monthus10phonon}
C. Monthus and T. Garel, 
Anderson localization of phonons in dimension d=1,2,3 : finite-size properties of the Inverse Participation Ratios of eigenstates,
Phys. Rev. B 81, 224208 (2010).

\bibitem{hastings03}
M. B. Hastings, 
Random Vibrational Networks and the Renormalization Group,
Phys. Rev. Lett. 90, 148702 (2003).

\bibitem{amir10}
A. Amir, Y. Oreg and Y. Imry, 
Localization, Anomalous Diffusion, and Slow Relaxations: A Random Distance Matrix Approach,
Phys. Rev. Lett. 105, 070601 (2010).

\bibitem{monthus10elastic}
C. Monthus and T. Garel, 
Random elastic networks : strong disorder renormalization approach,
J. Phys. A: Math. Theor. 44 085001 (2011).






\bibitem{strogatz00}
 S. H. Strogatz,
 From Kuramoto to Crawford:  exploring the onset of synchronization in populations of coupled oscillators,
Physica D 143, 1  (2000).

\bibitem{pikovsky01}
  A. Pikovsky, M. Rosenblum, and J. Kurths,
Synchronization: A Universal Concept in Nonlinear Science, 
Cambridge University Press, New York, 2001.

\bibitem{pikovsky15}
A. Pikovsky and M. Rosenblum,
Dynamics of globally coupled oscillators: progress and perspectives
Chaos 25, 097616 (2015).


\bibitem{kogan09}
O. Kogan, J. L. Rogers, M. C. Cross, and G. Refael, 
Renormalization group approach to oscillator synchronization,
Phys. Rev. E 80, 036206 (2009).

\bibitem{lee09}
T. E. Lee, G. Refael, M. C. Cross, O. Kogan and J. L. Rogers, 
Universality in the one-dimensional chain of phase-coupled oscillators,
Phys. Rev. E 80, 046210 (2009).












\bibitem{mohan10}
P. Mohan, R. Narayanan and T. Vojta,
Infinite randomness and quantum Griffiths effects in a classical system: the randomly layered Heisenberg magnet,
Phys. Rev. B 81, 144407 (2010)

\bibitem{monthus09wetting}
C. Monthus and T. Garel,
Random wetting transition on the Cayley tree: a disordered first-order transition with two correlation length exponents,
 J. Phys. A: Math. Theor. 42 165003 (2009).

\bibitem{monthus17dna}
C. Monthus, 
Strong Disorder Renewal Approach to DNA denaturation and wetting : typical and large deviation properties of the free energy,
J. Stat. Mech.  013301 (2017).


\bibitem{monthus14LRSGzero}
C. Monthus,
One-dimensional Ising spin-glass with power-law interaction : real-space renormalization at zero temperature,
 J. Stat. Mech. P06015 (2014).

\bibitem{monthus15fractal}
C. Monthus, 
Fractal dimension of spin glasses interfaces in dimensions d=2 and d=3 via strong disorder renormalization at zero temperature,
Fractals, 23, 1550042 (2015).


\bibitem{wang17}
W. Wang, M. A. Moore and H. G. Katzgraber, 
The Fractal Dimension of Interfaces in Edwards-Anderson and Long-range Ising Spin Glasses: Determining the Applicability of Different Theoretical Descriptions,
Phys. Rev. Lett. 119, 100602 (2017).

\bibitem{wang18}
W. Wang, M. A. Moore and H. G. Katzgraber, 
Fractal dimension of interfaces in Edwards-Anderson spin glasses for up to six space dimensions,
Phys. Rev. E 97, 032104  (2018).


\bibitem{schehr10}
G. Schehr and P. Le Doussal,
Extreme value statistics from the real space renormalization group: Brownian motion, Bessel processes and continuous time random walks,
J. Stat. Mech. (2010) P01009 (2010).

\bibitem{extreme}
G. Gy\"orgyi, N. R. Moloney, K. Ozog\'any, Z. R\'acz, and M. Droz,
Renormalization-group theory for finite-size scaling in extreme statistics
Phys. Rev. E 81, 041135 (2010);
E. Bertin and G. Gy\"orgyi,
Renormalization flow in extreme value statistics,
J. Stat. Mech. P08022 (2010);
I. Calvo, J. C. Cuchí, J. G. Esteve, and F. Falceto,
Extreme-value distributions and renormalization group,
Phys. Rev. E 86 041109 (2012);
F. Angeletti, E. Bertin, and P. Abry,
Renormalization flow for extreme value statistics of random variables raised to a varying power,
J. Phys. A: Math. Theor. 45 115004 (2012).

\bibitem{juhasz09extremal}
R. Juh\'asz,
A non-conserving coagulation model with extremal dynamics,
J. Stat. Mech.  P03033 (2009).



 \end{thebibliography}
\end{document}